\documentclass[12pt,preprint]{emulateapj}
\usepackage{graphicx}

\newcommand{\be}{\begin{equation}}
\newcommand{\bel}[1]{\be\label{eq:#1}}
\newcommand{\ee}{\end{equation}}
\newcommand{\ba}{\begin{eqnarray}}
\newcommand{\ea}{\end{eqnarray}}
\newcommand\bp{\begin{figure}}
\newcommand\ep{\end{figure}}
\newcommand\bpm{\begin{figure*}}
\newcommand\epm{\end{figure*}}
\newcommand{\btab}{\begin{tabular}}
\newcommand{\etab}{\end{tabular}}
\newcommand{\bt}{\begin{table}}
\newcommand{\et}{\end{table}}
\newcommand{\ben}{\begin{enumerate}}
\newcommand{\een}{\end{enumerate}}
\newcommand\reffig[1]{Figure \ref{fig:#1}}
\newcommand\refeq[1]{Equation \ref{eq:#1}}
\newcommand\refsec[1]{\S \ref{sec:#1}}
\newcommand\reftbl[1]{Table \ref{tbl:#1}}
\newcommand{\cm}{\rm{cm}}
\newcommand{\pc}{\rm{pc}}
\newcommand{\Ray}{\rm{R}}
\newcommand{\kjysr}{~\rm{kJy\,sr}^{-1}}

\newcommand{\bcn}{\begin{center}}
\newcommand{\ecn}{\end{center}}

\newcommand{\nup}{\nu_{\rm p}}
\newcommand{\hal}{H$\alpha$}

\begin{document}

\title{Identification of Spinning Dust in H$\alpha$--Correlated Microwave 
Emission}

\author{Gregory Dobler\altaffilmark{1,2} \& Douglas P. Finkbeiner\altaffilmark{1,3}}

\altaffiltext{1}{
Institute for Theory and Computation, 
Harvard-Smithsonian Center for Astrophysics, 60 Garden Street, MS-51,
Cambridge, MA 02138 USA
}
\altaffiltext{2}{gdobler@cfa.harvard.edu}
\altaffiltext{3}{dfinkbeiner@cfa.harvard.edu}

\begin{abstract}
CMB experiments commonly use maps of H$\alpha$ intensity as a spatial
template for Galactic free--free emission, assuming a power law
$I_{\nu} \propto \nu^{-0.15}$ for the spectrum.  Any departure from
the assumed free--free spectrum could have a detrimental effect on
determination of the primary CMB anisotropy.  We show that the
H$\alpha$-correlated emission spectrum in the diffuse WIM is
\emph{not} the expected free--free spectrum at WMAP frequencies.
Instead, there is a broad bump in the spectrum at $\sim$ 50 GHz which
is consistent with emission from spinning dust grains.  Spectra from
both the full sky and smaller regions of interest are well fit
by a superposition of a free--free and ``warm ionized medium'' Draine
\& Lazarian (1998) spinning dust model, shifted in frequency.  
The spinning dust emission is $\sim$
5 times weaker than the free--free component at 50 GHz, with the null
hypothesis that the H$\alpha$-correlated spectrum is pure free--free,
ruled out at $\geq
8\sigma$ in all regions and $>100\sigma$ for the full sky fit. 
\end{abstract}

\section{Introduction}

During the course of measuring the cosmic microwave background (CMB)
anisotropy, the \emph{Wilkinson Microwave Anisotropy Probe} (WMAP) has
produced the most detailed and sensitive maps of interstellar medium
(ISM) emission between 20 and 100 GHz taken to date.  The quality of
the data has profoundly influenced our understanding of the four main
ISM continuum emission mechanisms in this frequency range
\citep{bennett03,finkbeiner03,hinshaw07,DF07}.  At the highest WMAP
frequencies, the Galaxy is dominated by thermal dust emission, traced
by far IR maps \citep{finkbeiner99}.  At lower
frequencies, synchrotron radiation from supernova shock accelerated
electrons spiraling in the Galactic magnetic field makes up the
majority of the emission.  In the inner Galaxy, the synchrotron
emission is brighter relative to low-frequency maps, and has a harder
spectrum.\footnote{\citet{finkbeiner04} has previously interpreted this as a 
separate component, the synchrotron ``haze'', but this interpretation is
controversial \citep{finkbeiner04,DF07}.  This haze component is
spatially orthogonal to the other templates by construction, and
though we include it in our fits below, its existence has negligible
effect on the \hal-correlated spectrum.}  Next, free--free emission
from unbound electrons interacting with ions in $\sim 10^4$K ionized
gas has a harder spectrum than synchrotron, and is detectable in all bands.
Finally, there is a bump in the spectrum of dust-correlated emission
at $\sim 20$ GHz, sometimes called ``Foreground X'' \citep{deO02}.
Over the last decade, it has been recognized that this could be
electric dipole emission from rapidly rotating dust grains \cite[``spinning
dust'';][]{DL98b}.

\subsection{Foreground X: spinning dust?}
Spinning dust is the electric dipole emission from the smallest dust
grains which have a non-zero electric dipole moment and are spun
up by several different interactions with the interstellar medium
\citep[e.g., ion collisions, the interstellar radiation field, etc.; see][]{DL98b}.  
Spinning dust was considered as a potential source of
radio and microwave emission in other contexts \citep{erickson57,ferrara94}.
  Because the model depends on the grain size
distribution, ionization fraction, electric dipole moments, etc., the
peak frequency can be anywhere from $10-50$ GHz, but is usually in the
$20-40$ GHz range.  This
dependence of the spectrum on the ISM parameters has made spinning
dust difficult to definitively identify, but may provide the qualitative
behaviors necessary to make a fool-proof detection.

In the COBE/DMR data, \citet{kogut96} found that the dust-correlated
emission does not simply fall off monotonically with decreasing
frequency as would be expected for a thermal dust tail, but instead was greater at 
31 GHz than at 53 GHz.  This behavior
was confirmed by the Saskatoon CMB experiment \citep{deO97},
the 14.5 and 32 GHz OVRO data \citep{leitch97},
the Cottingham \& Boughn 19.2 GHz survey \citep{deO98}, 
the Tenerife 10 \& 15 GHz survey \citep{deO99}, and the
QMAP 30 GHz data \citep{deO00}. 
A small survey of dust clouds with the Green Bank 42m telescope at 5,
8, \& 10 GHz \citep{F02} revealed two candidates for spinning dust
clouds: L1622, later
confirmed \citep{finkbeiner04}; and LPH~201.663+1.643, which turned
out to be incorrect \citep{dickinson06}. 

Given this evidence, it was already clear prior to the WMAP data
release in early 2003, that there was some non-standard emission
mechanism, but it was unclear whether this mechanism was spinning dust
\citep{DL98b} or some other process.

WMAP \citep{bennett03} found that the dust-correlated emission falls
from 94 to 61 GHz and then rises steadily to 23 GHz.  The fact that
synchrotron typically dominates at lower frequencies led \citet{bennett03} to
interpret the low frequency rise as ``dust-correlated synchrotron''.
The reasoning was that because dusty regions correlate with regions of
star formation activity, electrons in the vicinity could be
accelerated by supernova shocks leading to synchrotron emission.
\citet{finkbeiner04} pointed out that WMAP is also consistent with
\citet{DL98b} if their model could be shifted down in frequency a bit.

Because the WMAP frequency coverage ($23-94$ GHz) coincides with the
range where the spinning dust and synchrotron spectra are similar,
data at lower frequencies, where the spectra diverge, have been used to
test the dust-correlated synchrotron hypothesis.  The WMAP ``synchrotron''
template has a much weaker cross-correlation at $10-15$ GHz than
expected for actual synchrotron \citep{deO04}.  Also, 8 \& 14 GHz data from
the Green Bank Galactic Plane Survey \citep{langston00} show a rising
spectrum consistent with spinning dust but not synchrotron
\citep{gb04} and implying that the majority of dust-correlated
emission at 23 GHz is Foreground X.  Comparing the Cottingham \&
Boughn survey to WMAP also shows a rise from 19 to 23 GHz
\citep{boughn07}.  The polarization properties \citep{page07,hinshaw07} in the
3-yr WMAP data also favor the spinning dust hypothesis.

In spite of all this evidence, there are weaknesses in each of these
arguments.  Some use only low-latitude data and may not apply to the
diffuse ISM \citep{gb04}.  Others apply to only one cloud \citep{F02},
have poor signal to noise \citep{deO04}, or a short lever arm in
frequency \citep{boughn07}.  However, taken together, they are very
strong evidence against the dust-correlated synchrotron hypothesis,
but they are not specifically evidence \emph{for} spinning dust.
Other mechanisms such as magnetic dipole dust emission, caused by the
thermal fluctuation of the magnetization in the dust grains
\citep{DL99}, could conceivably explain all these results.  This
ambiguity motivated our search for further clues in the WMAP 3-yr
data.

\subsection{WMAP Multilinear Regression}
In \citet[hereafter, DF07]{DF07} we describe our method for
determining the spectral shape of each of the four primary WMAP
foreground components: free--free, dust, soft synchrotron, and hard synchrotron 
\citep[termed the ``haze'' by][]{finkbeiner04} emission.  We model the WMAP data 
as a linear combination of the CMB
and the 4 foregrounds, each described by a specific spatial template.
For free--free emission
we use the \hal\ map described in \refsec{halpha-em}, for dust (thermal
and spinning) emission we use the \citet{schlegel98} dust map
evaluated at 94 GHz by \citet{finkbeiner99}, for soft synchrotron we
use the \citet{haslam82} 408 MHz map, and for the haze we use a simple
$1/r$ template where $r$ is the distance from the Galactic center.
The individual WMAP bands were completely decoupled in our fit, and we
assumed \emph{nothing} about the spectrum of each foreground, only
that its morphology was traced by the template in all bands.  We
performed our fits over both the (nearly) full sky and also on smaller
regions of interest.  Point sources and regions where dust extinction
makes the \hal\ map unreliable (where $A(\mbox{\hal}) =
2.65\,E(B-V) > 1\mbox{ mag}$) were masked out from the fits.

We showed that the spectrum of \hal-correlated emission is
\emph{not} the usual free--free spectrum.  Rather, there is
a bump around $\nu \sim 50$ GHz.  In \refsec{sdust-em} we explain why
the \hal\ map should actually trace spinning dust excited by
collisions with ions.  In \refsec{spectra} we present \hal-correlated
emission spectra for both the (nearly) full sky as well as smaller
regions of interest and show that, indeed, the spectra are well
characterized as a superposition of free--free and spinning dust type
spectra.  We summarize this work and discuss implications for future
observations in \refsec{discussion}.

\section{Emission measure as a tracer of spinning dust}

In this section we argue that in fully ionized environments, the emissivity 
density ($j_\nu$) of spinning dust emission is proportional to dust density times 
some excitation function related to the ion density.  The emission comes from the 
very smallest grains, which are spun up by a collision with an ion and spin down 
as they radiate.  In the limit where the spin down time is small compared to the 
time between collisions, the rotation is episodic and the emissivity also scales 
linearly with the ion number density.  In \S \ref{sec:spinem}, we assume the episodic limit in order to build some intuition about how WIM-correlated spinning dust might behave.

\subsection{Free--free emission measure}
\label{sec:halpha-em}

The map assembled by \citet{finkbeiner03} from the VTSS
\citep{dennison98}, SHASSA \citep{gaustad01}, and WHAM
\citep{haffner03} surveys, traces the \hal\ recombination line
emission from hot interstellar gas.  At WMAP frequencies, this \hal\
map\footnote{available at \texttt{http://www.skymaps.info}} is used to
estimate the thermal bremsstrahlung, or free--free emission, resulting
from the interaction of the electrons and protons from ionized
hydrogen \citep{bennett03,finkbeiner04,hinshaw07}.  The \hal\ map is
corrected for dust absorption using the prescription given by Finkbeiner
(2003; but see Dickinson et al. 2003 for an alternative).  In this
frequency range the free--free specific intensity $I_\nu$ (in Jy/sr) is
often approximated as a power law, \bel{ff-spec} I_{\nu} \propto
\nu^{\alpha}, \ee with $\alpha \approx -0.15$.  Although this is a
good approximation, in our analysis we use the expression given by
\citet{spitzer} for the free--free spectrum.

Recombination line emission is a two particle process (involving the
capture of free electrons by a proton nucleus) and so the emissivity
is proportional to $n_e n_i$, which for fully ionized H is simply the
square of the number density of gas particles, $j_{\mbox{\hal}}\propto
n^2$.  Therefore, the \hal\ intensity corresponds to an \emph{emission
  measure} (EM) defined as $E_m \equiv \int n^2 dl$ where the integral
is taken along the line of sight.  The \hal\ intensity is usually
measured in Rayleighs (1 R $\equiv 10^6/4\pi$ photons s$^{-1}$
cm$^{-2}$ sr$^{-1}$) with an EM of 1 cm$^{-6}$pc $\approx 0.6$ R for
gas at $\sim 8000$ K.

\subsection{Spinning dust emission measure}
\label{sec:spinem}

\citet[hereafter, DL98]{DL98b} present a model for spinning dust
emission in which tiny dust grains with non-negligible electric dipole
moment are spun up by ion collisions; polycyclic aromatic hydrocarbons (PAHs) 
generally have a non-zero dipole, because perfect symmetry is rare in large 
molecules.  DL98 estimated the astronomical value of this parameter (actually the 
electric dipole moment per root mass) from measured electric dipole moments for 
dozens of known PAH molecules.  In the limit where ion interactions provide the 
dominant spin-up mechanism (see DL98 for details) and assuming full ionization, 
constant gas/dust ratio, and the same grain size distribution everywhere, the 
resulting \emph{total} spinning dust emission is $I \equiv \int I_{\nu} d\nu 
\propto \int n_i n_d dl \propto \int n_i^2$.  
With the additional assumption that the emission is episodic, the spectrum of the emission does not vary with $n_i$, but the amplitude does. 
That is, the spinning dust emission 
should look like an EM, and be traced by \hal.  This approximation is relevant 
where the collisional excitation is primarily due to ions (at least for the smallest grains), e.g. for the ``warm ionized medium'' (WIM) parameters used by DL98.

The WIM-correlated emission claimed here augments the ``usual''
spinning dust emission in the ``cold neutral medium'' (CNM) which has
a temperature $T_{\rm CNM} \sim 100$ K and is traced by the
\citet{schlegel98} dust map evaluated at 94 GHz by
\citet{finkbeiner99} (FSD99).

\subsection{Estimating the specific intensity}
\label{sec:sdust-em}

For a spherical dust grain of size $a$, rotating with angular frequency $\omega$, 
and with a dipole moment $p_0$, the rotational energy is,
\bel{rotenergy}
  E = \frac{1}{2} I \omega^2 = \frac{1}{5} m a^2 \omega^2,
\ee
where $m$ is the grain mass.  The energy loss rate is the total power emitted by 
the grain,
\be
  \dot{E} = \frac{1}{12\pi}\frac{\mu_0}{c} p_0^2 \omega^4
\ee
and the spin down time is,
\be
  \tau_s \equiv E/\dot{E} \propto a^2 p_0^{-2} \omega^{-2}.
\ee
The time between collisions is just,
\be
  \tau_c \sim \lambda/v \propto a^{-2} n_H^{-1}
\ee
where $v$ is the thermal velocity of the ion and $\lambda=(\pi a^2
n_H)^{-1}$ is the mean free path.

In the limit where the spin down time is much \emph{smaller} than the time between 
collisions ($r \equiv \tau_s/\tau_c \ll 1$), the excitation is episodic.  For the 
WIM parameters in DL98, this ratio is in fact $\sim 1$; however, we point out that 
\be
  r = \tau_s/\tau_c \propto a^4 n_H p_0^{-2} \omega^{-2}
\ee
so that very small changes to the grain size distribution can substantially affect 
the spin down time compared to the time between collisions.  In fact, if each ion 
collision imparts equal energy, then from \refeq{rotenergy}, $\omega \propto 
a^{-2}$.  Since $p_0 \propto N^{1/2} \propto a^{3/2}$ (see DL98) 
then in this case, $1/r$ diverges as $a^{-5}$.  The bottom line is that the 
episodic approximation is sensitively dependent on the ion density, the dipole 
moment, and 
\emph{especially} the grain size distribution. For the purposes of estimating a 
specific intensity, we will now assume the episodic approximation with the caveat 
that it may break down in dense environments, \citep[e.g., HII 
regions.  In fact,][find no evidence for spinning dust at 14.2 and 17.9 
GHZ in 16 Galactic HII regions]{scaife07} or for very small dipole moments.  
We emphasize that this does not affect the fact the the \emph{total intensity} $I$ 
should be proportional to $\int n_d n_H d\ell$ so that spinning dust emission 
should be traced by an EM map.

Draine \& Lazarian evaluate $j_\nu/n_H$ for parameters representative
of various interstellar environments.  In the limit that $j_\nu/n_H$
is a constant with $n_H$, then it equals $I_\nu/N(H)$.  However, in
our case it is not.  As a reference model, we take the DL98 WIM model
evaluated for $n_H=0.1~\cm^{-3}$ which peaks near 25 GHz with a
peak emissivity\footnote{The units of the ordinate, $j_\nu/n_H$, 
of Figs. 9-11 in DL98 is incorrectly labeled (Jy sr$^{-1}$) 
rather than (Jy cm$^2$\,sr$^{-1}$), an error propagated by Finkbeiner
et al. (2002).}
of $j_{p, {\rm ref}}/n_H = 5\times 10^{-21} \kjysr \cm^2$
or $j_{p, {\rm ref}} = 5\times10^{-22} \kjysr \cm^{-1}$. 
Multiplying the dependence on dust and gas density we have\footnote{
Again, the dependence on $n_H$ in \refeq{episod} is only strictly correct in the 
episodic limit, though in reality the deviation is likely small.
}
\bel{episod}
j_{p} = j_{p, {\rm ref}} \left(\frac{n_H}{n_{H,{\rm ref}}}\right)
\left(\frac{n_d}{n_{d,{\rm ref}}}\right)
\ee
where $n_{H,{\rm ref}}$ and $n_{d,{\rm ref}}$ are the density of H and small dust
grains, respectively, used in the reference model.  For full
ionization of H ($n_H = n_i$) and constant gas/dust ratio
($n_H/n_{H,{\rm ref}} = n_d/n_{d,{\rm ref}}$) we have
\begin{eqnarray}
j_{p} = j_{p, {\rm ref}} \left(\frac{n_H}{0.1\,\cm^{-3}}\right)^2 \\
=5\times10^{-20} n_H^2 \kjysr \cm^5 \\
=0.15 n_H^2 \kjysr \cm^6\pc^{-1}
\end{eqnarray}
and the peak specific intensity is then 
\begin{equation}
I_{p}=\int j_{p} dl = 0.15 \kjysr \left(\frac{E_m}{1\,\cm^{-6}\pc}\right)
\end{equation}
or
\begin{equation}
I_{\rm p}/I_{\mbox{\hal}} = 0.09\,\kjysr\,\Ray^{-1},
\end{equation}
where $I_{\mbox{\hal}}$ is measured in R.  For comparison, the ratio
of free--free to \hal\ for a temperature of $T_e\sim 8000$K at 41 GHz
(WMAP Q-band) is $I_{41}/I_{\mbox{\hal}} \sim
0.15\,\kjysr\,\Ray^{-1}$.

Although this is a rough estimate, it suggests that the intensity of
the free--free emission at 41 GHz and the peak of the spinning dust
emission should be of the same order of magnitude.
Throughout this discussion we make the assumption that the DL98 WIM
spectrum can be shifted by a factor of 2 in frequency.  A more
thorough analysis would require that the DL98 model be evaluated for
some set of parameters (ion density, electric dipole moment per root
mass, etc.) that give the actual spectrum observed.  For our purposes
this naive shift is adequate. 

\bp
\begin{center}
  \includegraphics[width=0.375\textwidth]{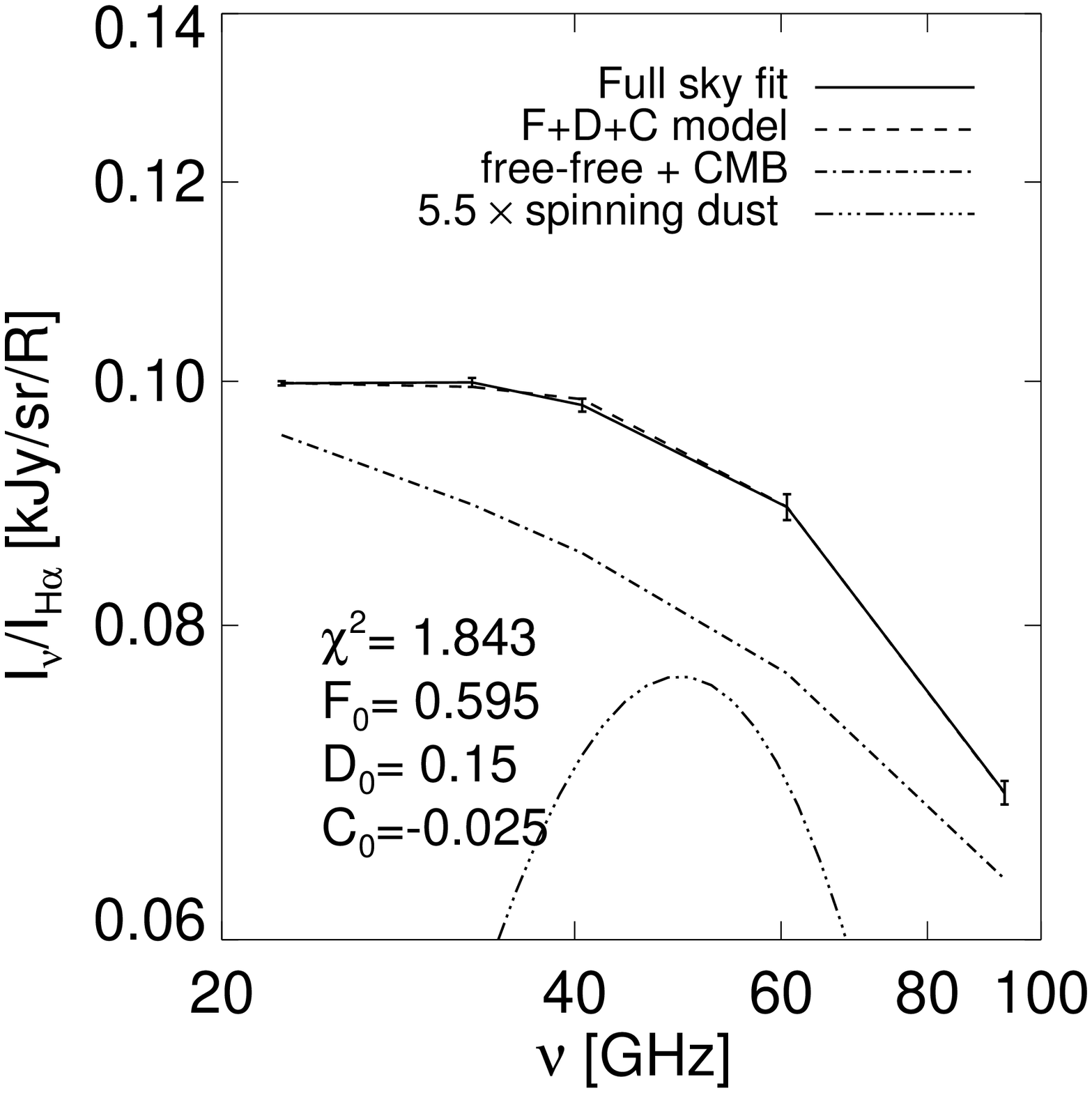}
  \includegraphics[width=0.375\textwidth]{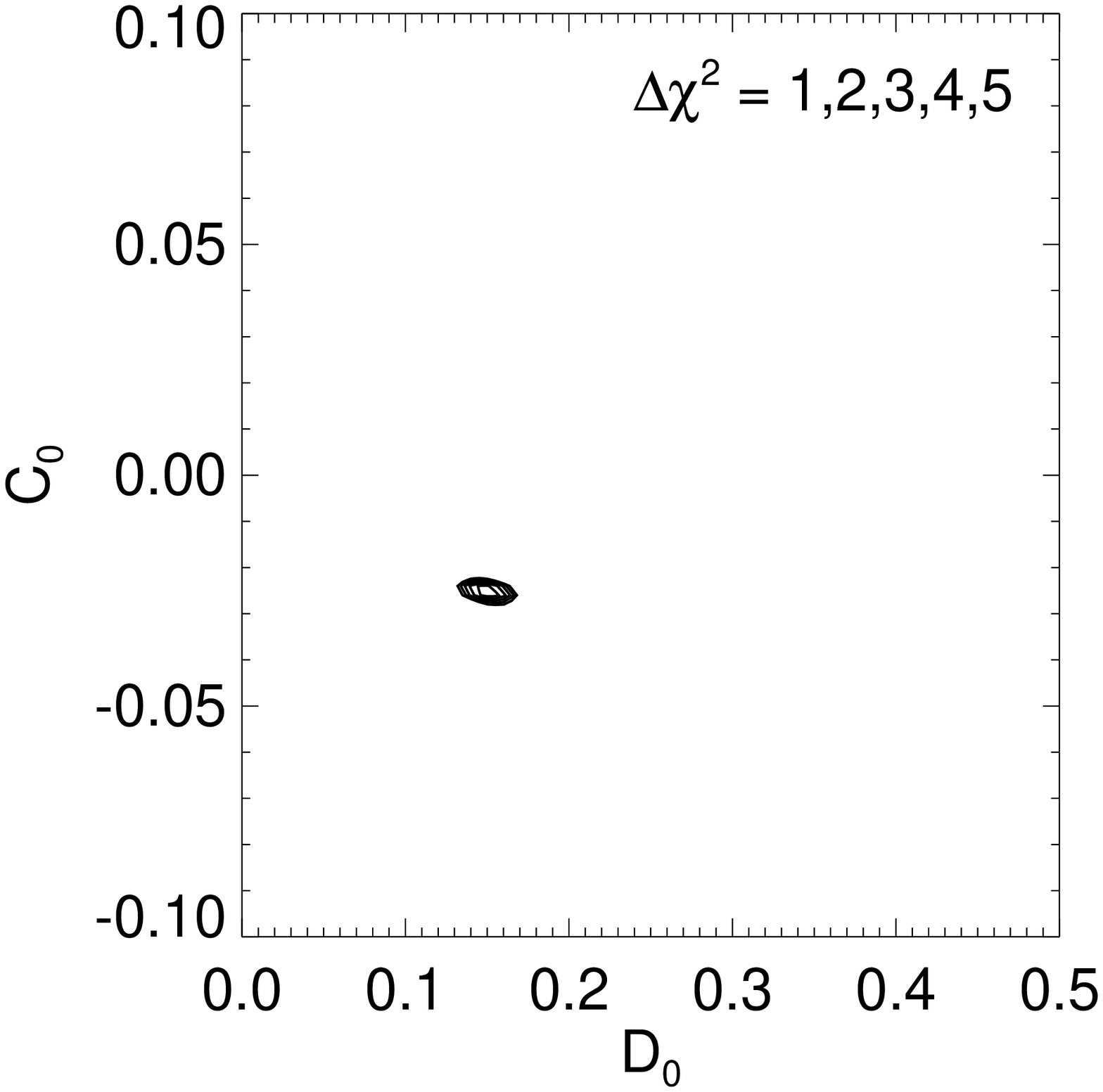}
  \includegraphics[width=0.375\textwidth]{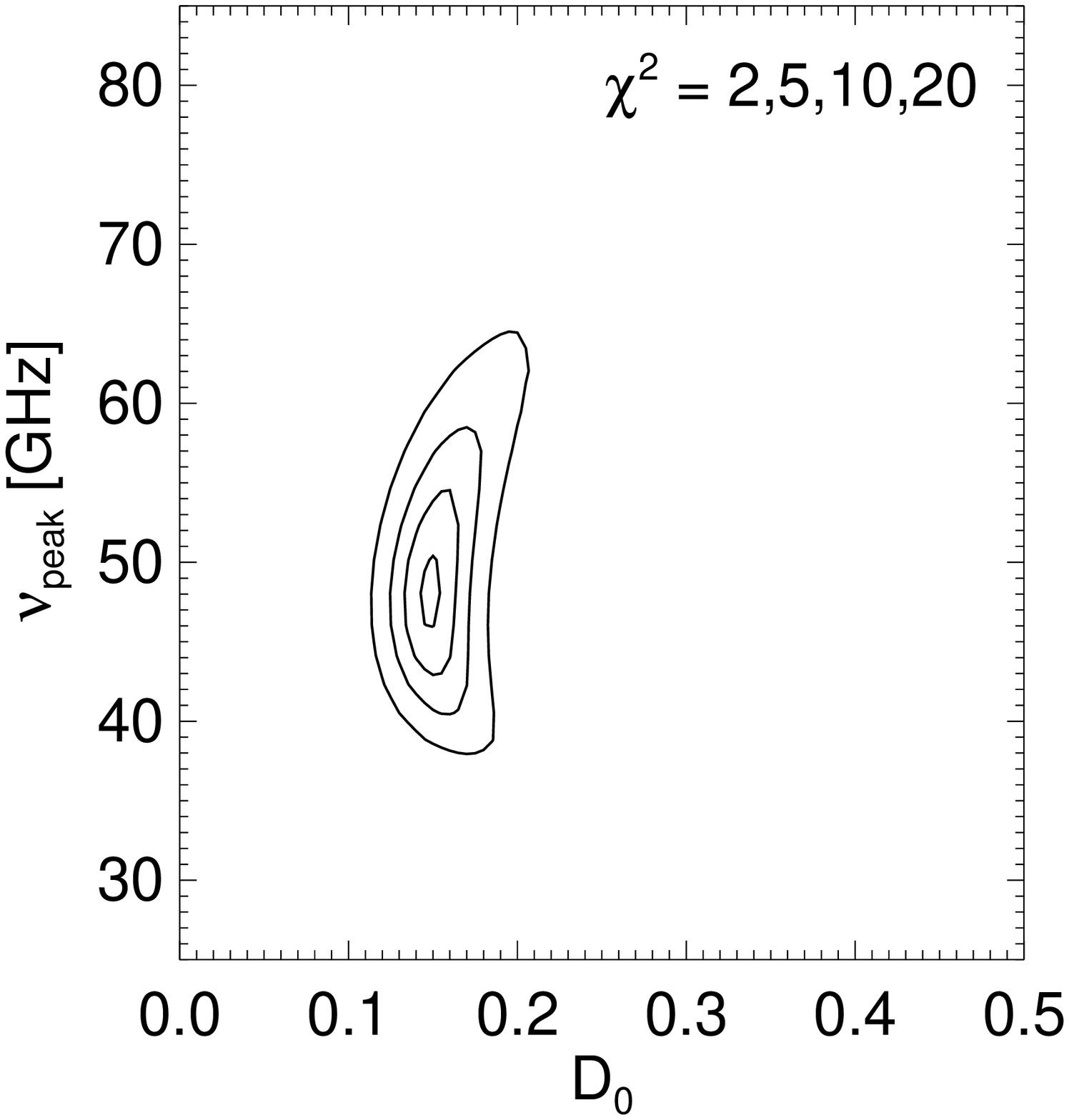}
\caption{
\emph{Top panel:} The best $F+D+C$ (free--free plus spinning dust plus CMB) 
model fit to the \hal-correlated emission over the full sky.  The peak frequency 
is fixed to $\nu_{\rm peak} = 50$ GHz.  This two component plus CMB 
contamination model works surprisingly well with a spinning dust component that 
is 6.23 times weaker than the free--free (see \reftbl{fit-res}).  \emph{Middle 
panel:} $\Delta\chi^2$ contours for the best fit free--free coefficient in the 
spinning dust/CMB plane.  The null hypothesis that the \hal-correlated emission 
can be well fit by only a $F+C$ model (i.e., $D_0 = 0$) is ruled at very high 
confidence ($335\sigma$).  \emph{Bottom panel:} $\chi^2$ contours in the 
spinning dust/$\nu_{\rm peak}$ plane.  Though the peak frequency is not very 
well constrained, we can rule out $\nu_{\rm peak} < 40$ GHz and $\nu_{\rm peak} 
> 60$ GHz at $>4\sigma$ confidence.
}\label{fig:hdc-fs}
\end{center}
\ep

\bpm
\centerline{
  \includegraphics[width=0.23\textwidth]{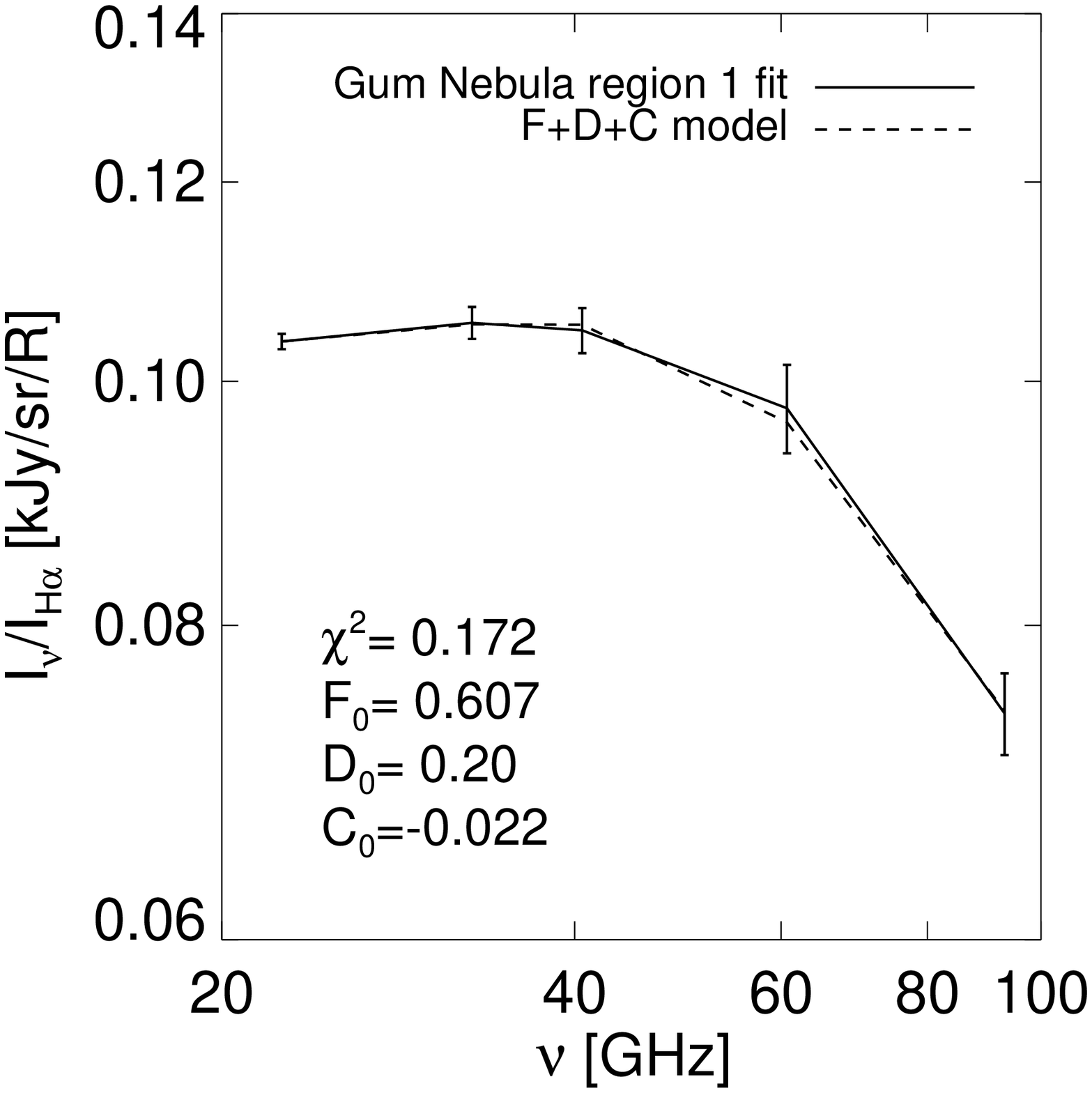}
  \includegraphics[width=0.23\textwidth]{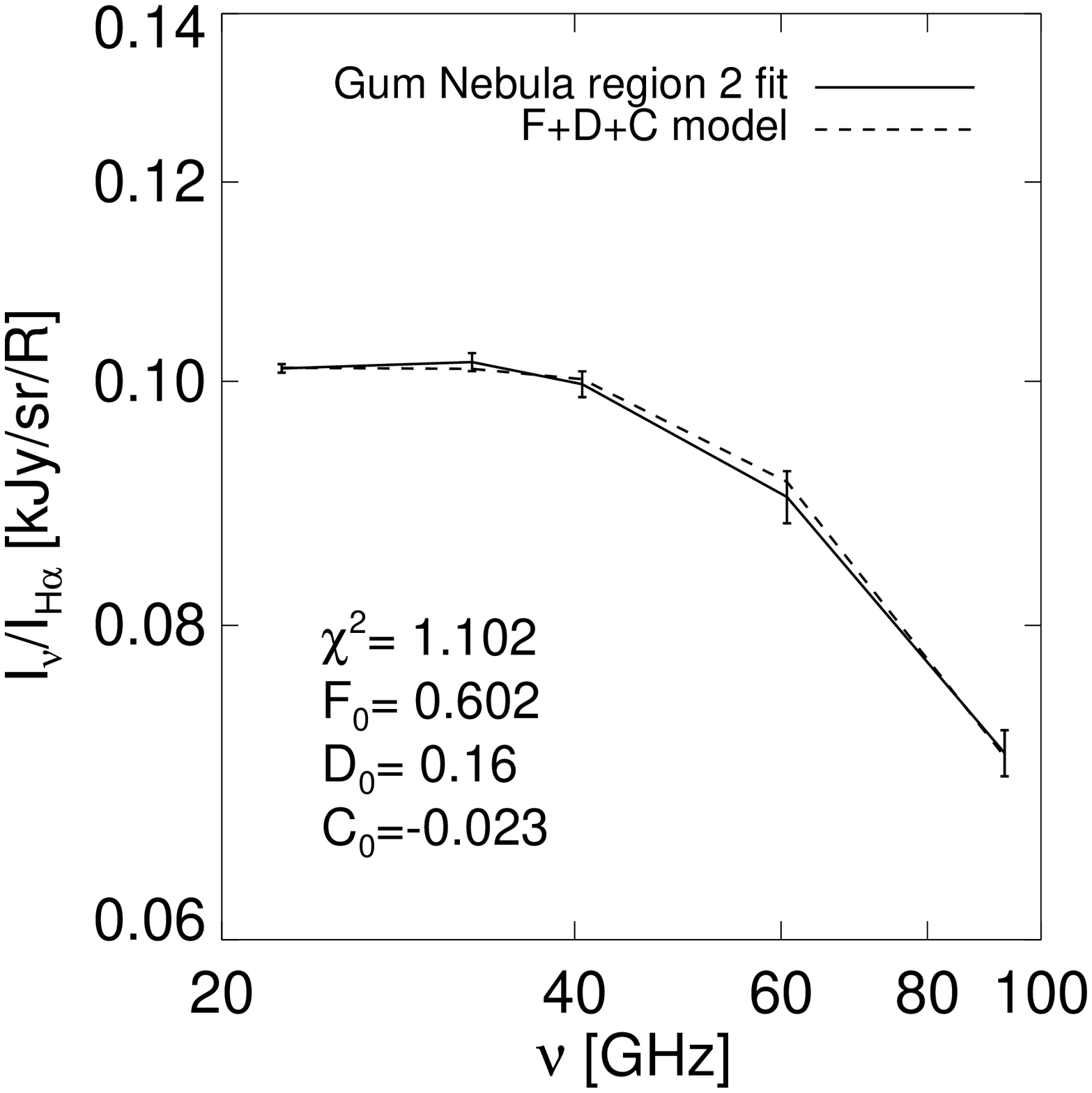}
  \includegraphics[width=0.23\textwidth]{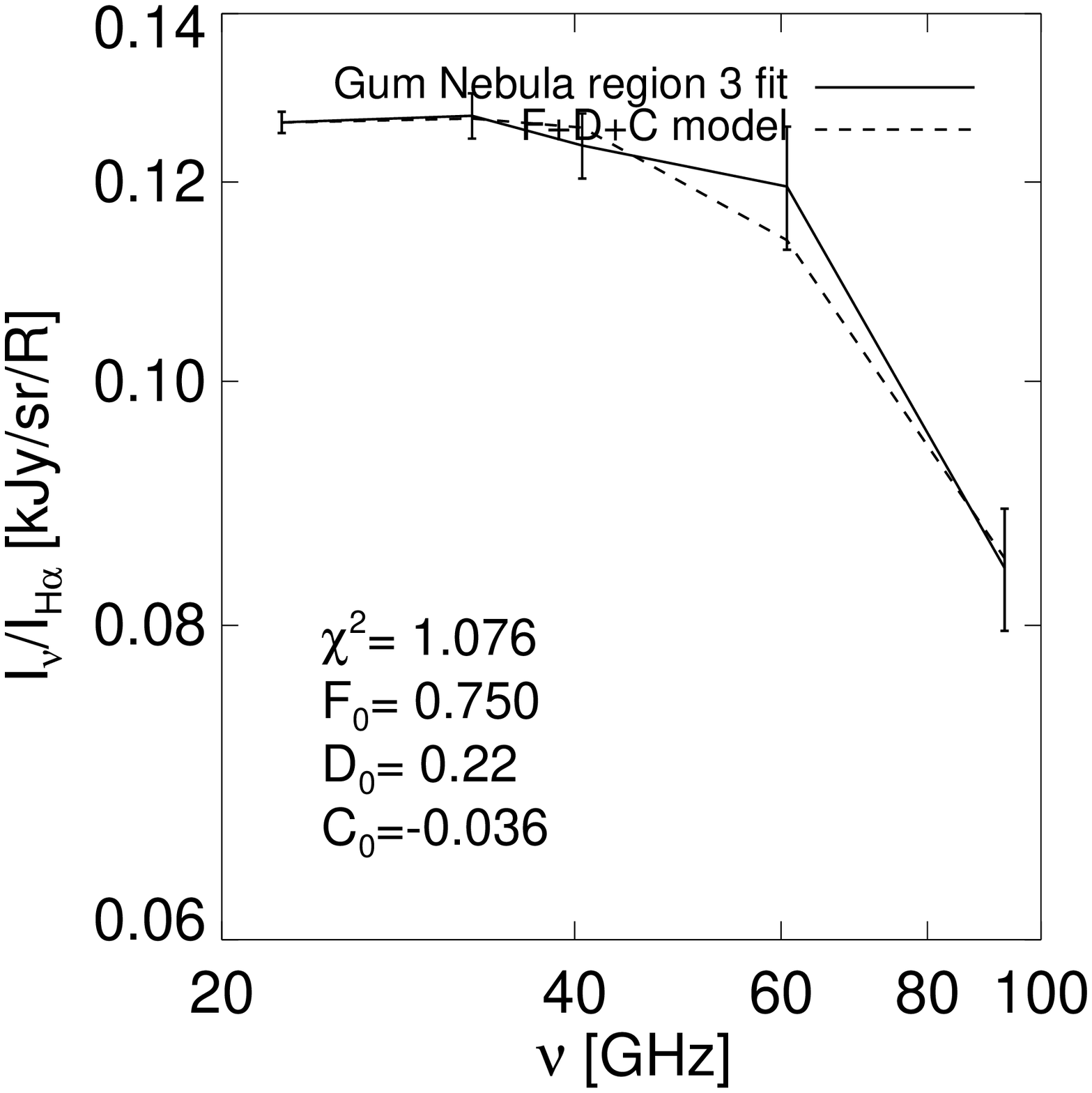}
  \includegraphics[width=0.23\textwidth]{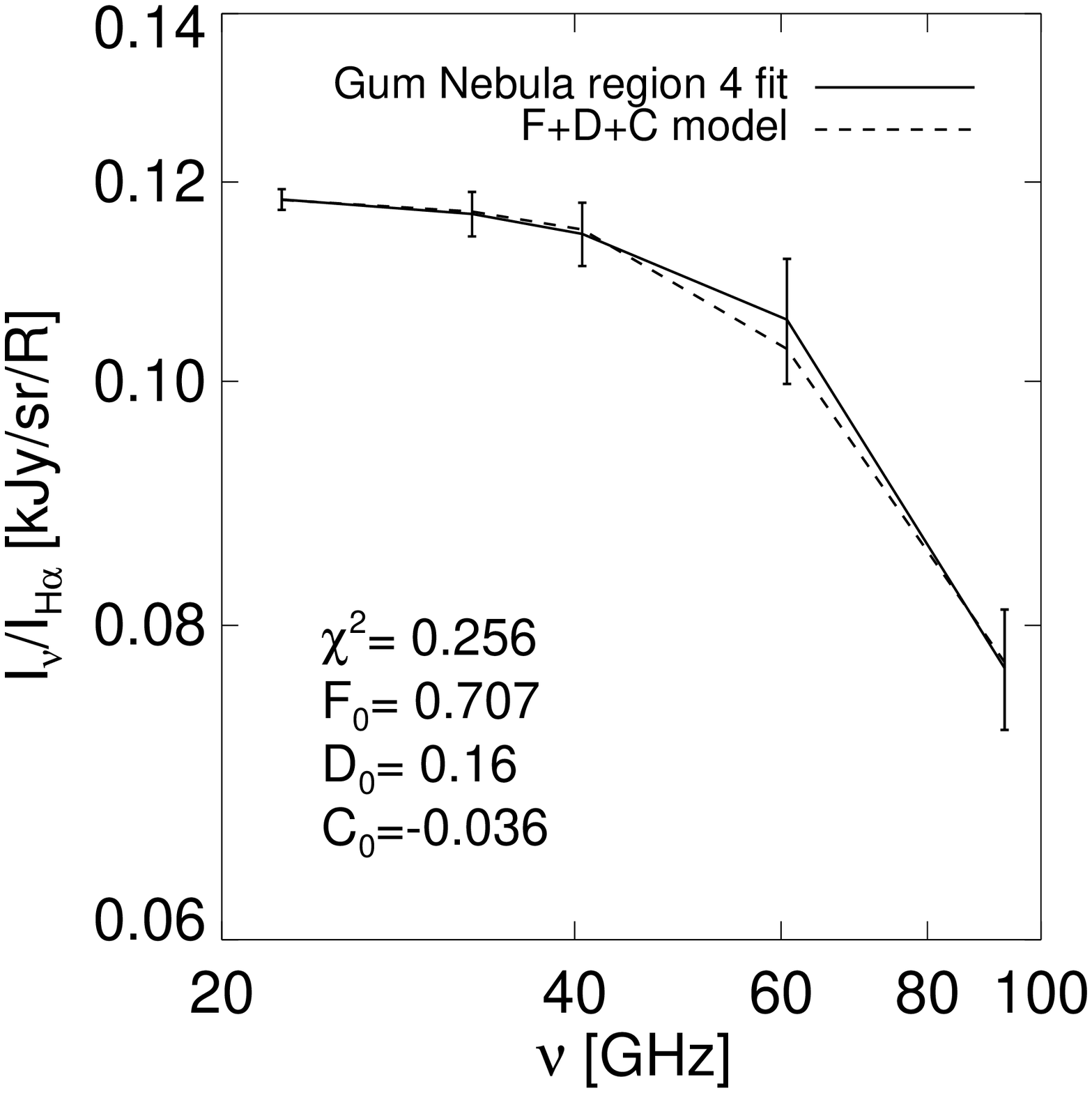}
}
\centerline{
  \includegraphics[width=0.23\textwidth]{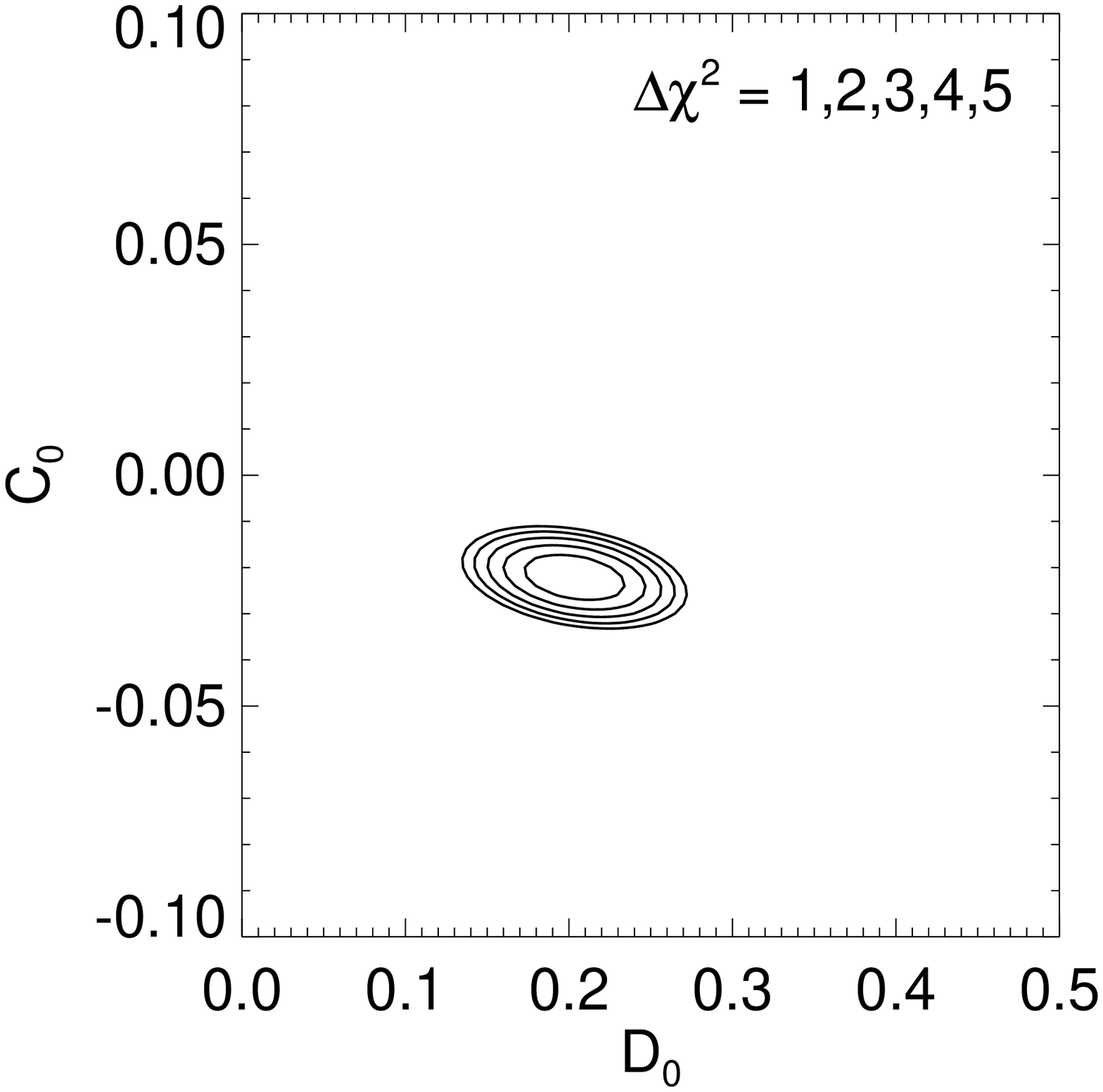}
  \includegraphics[width=0.23\textwidth]{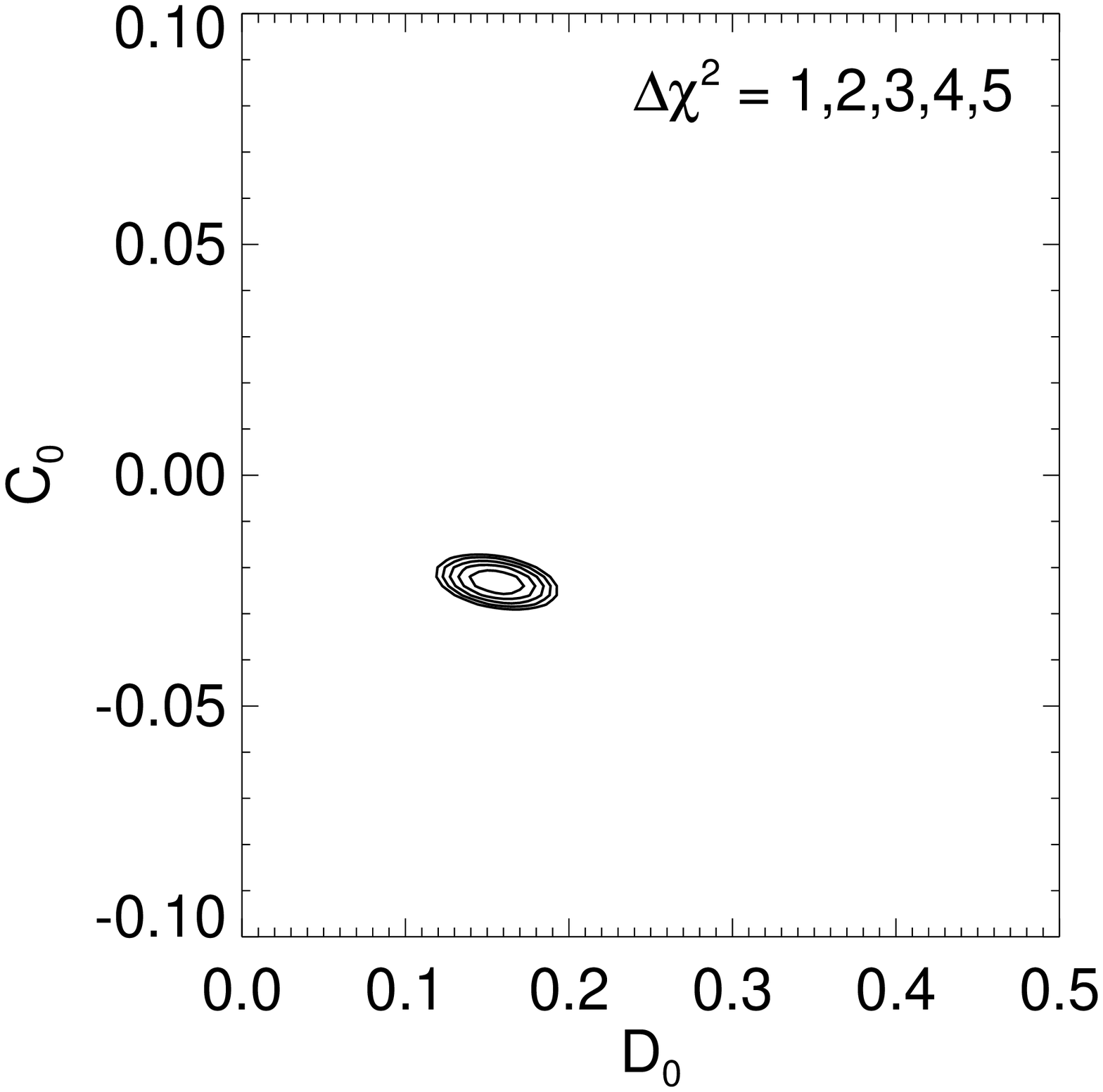}
  \includegraphics[width=0.23\textwidth]{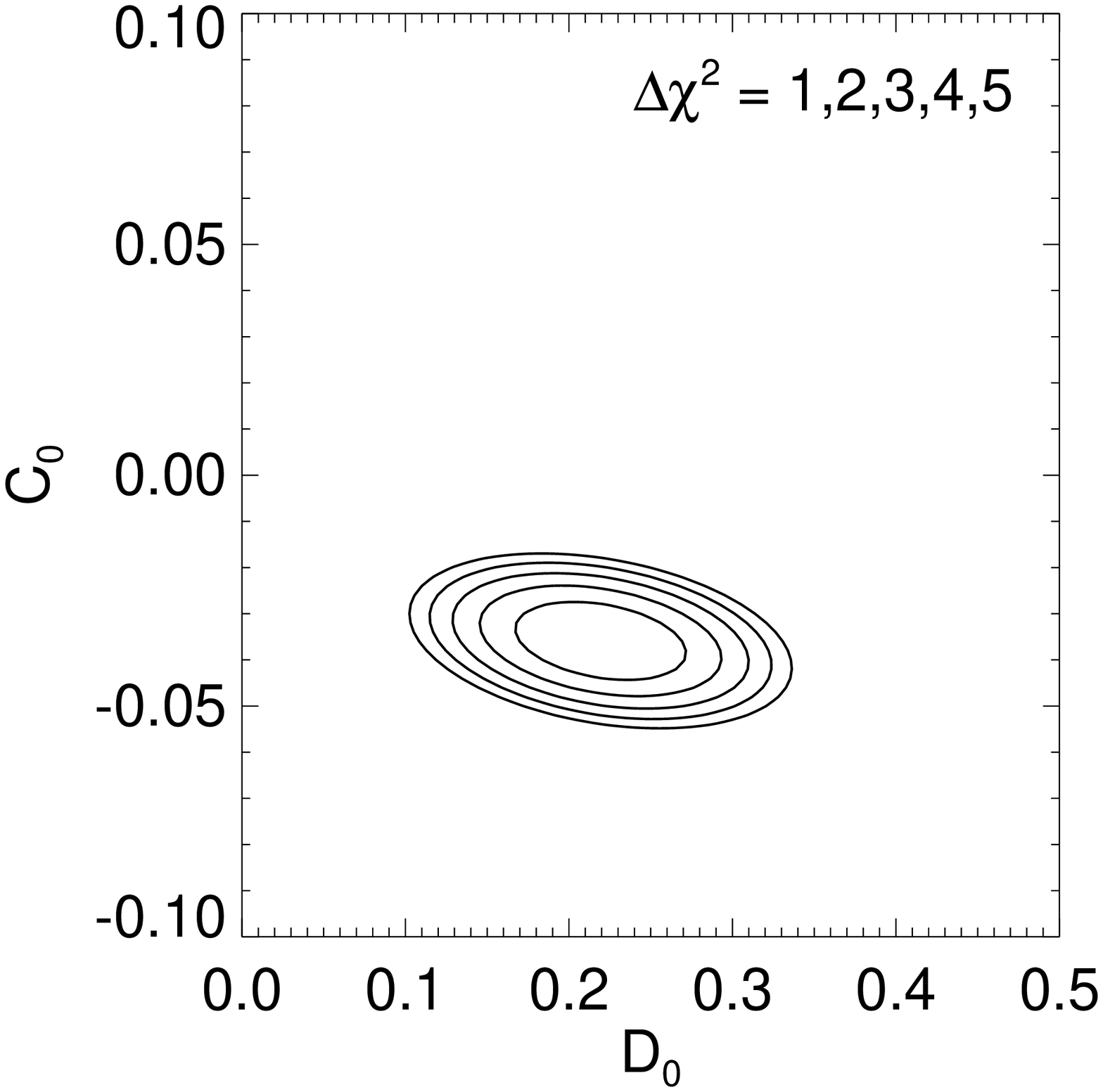}
  \includegraphics[width=0.23\textwidth]{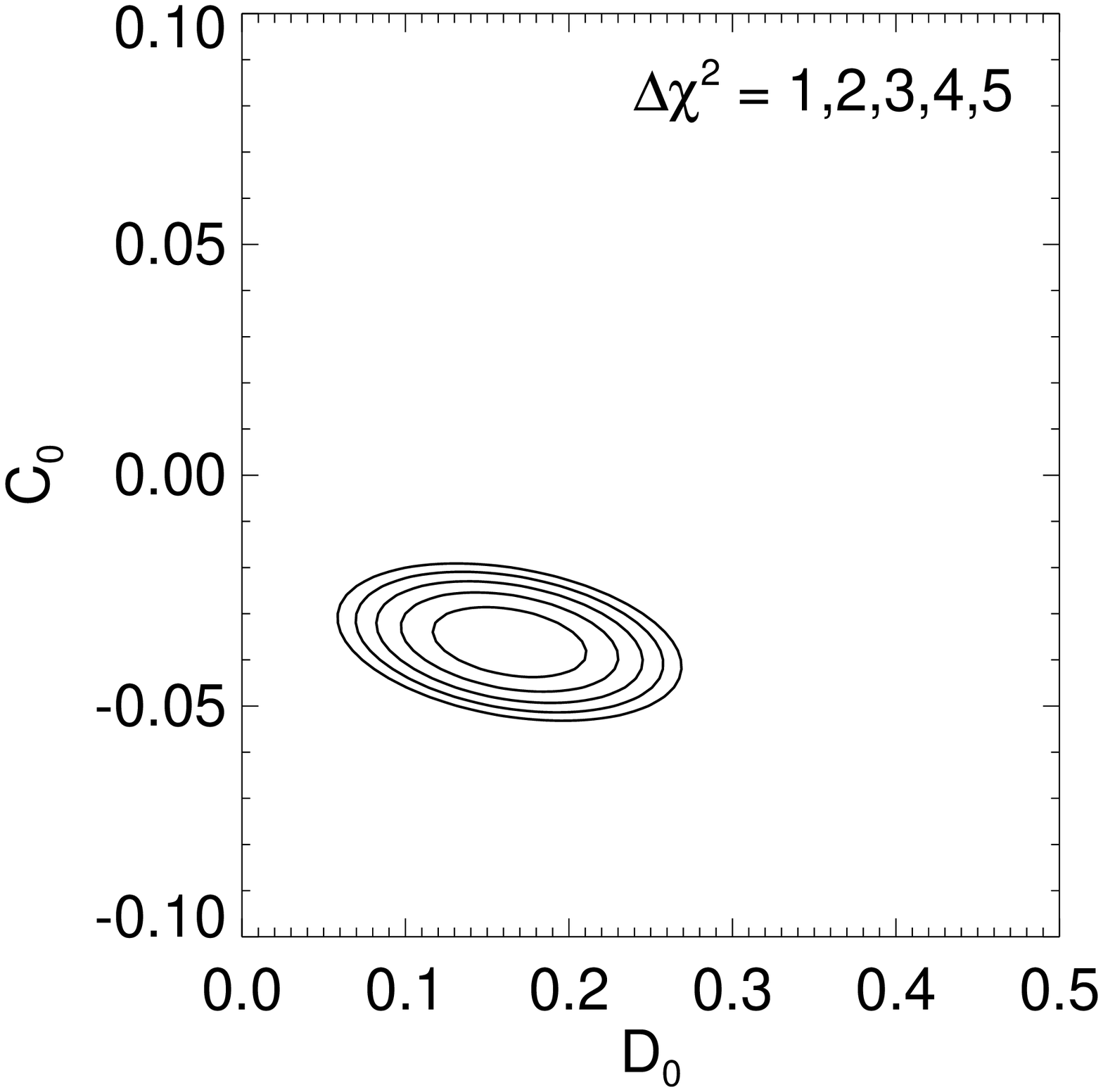}
}
\centerline{
  \includegraphics[width=0.23\textwidth]{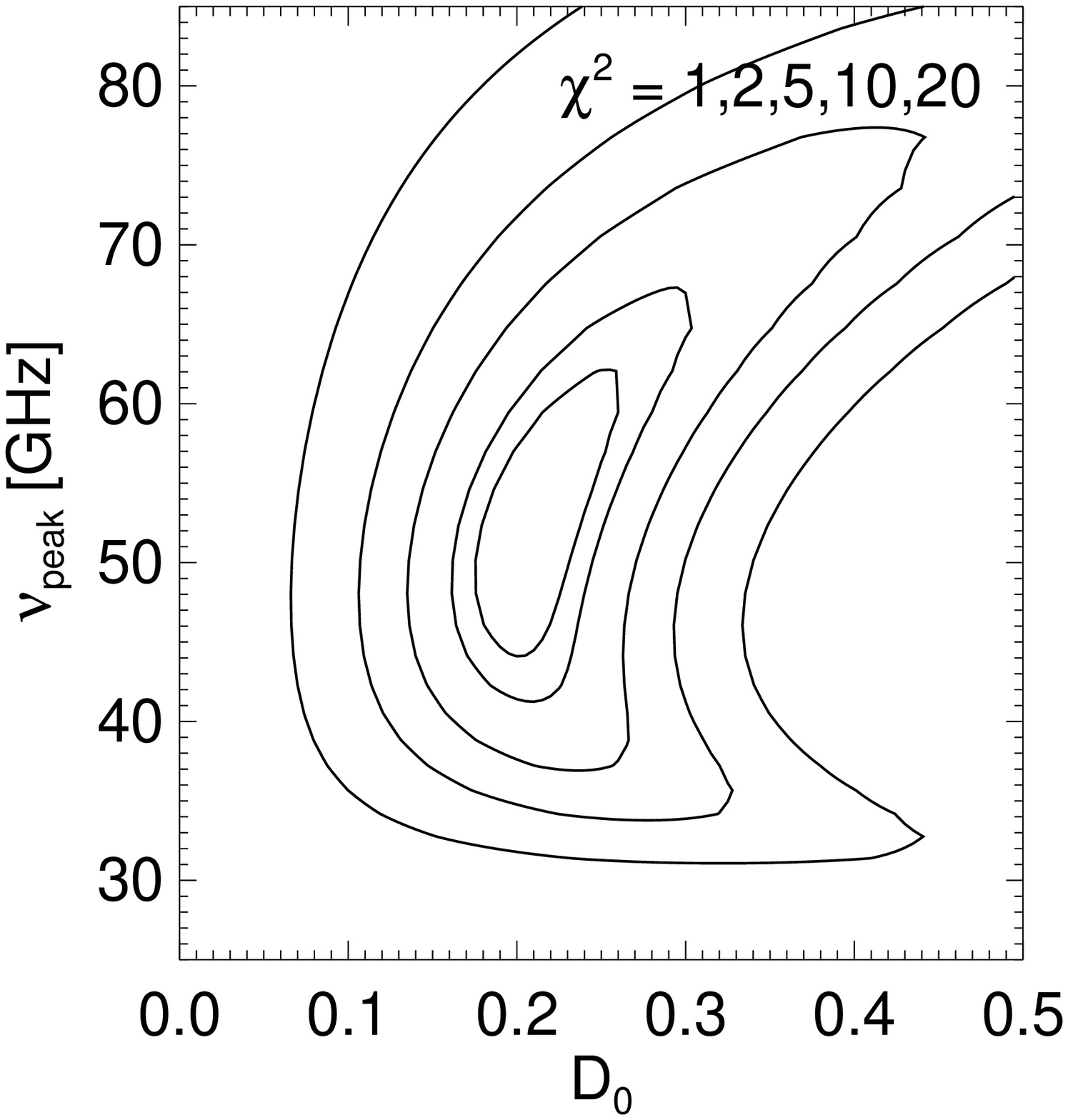}
  \includegraphics[width=0.23\textwidth]{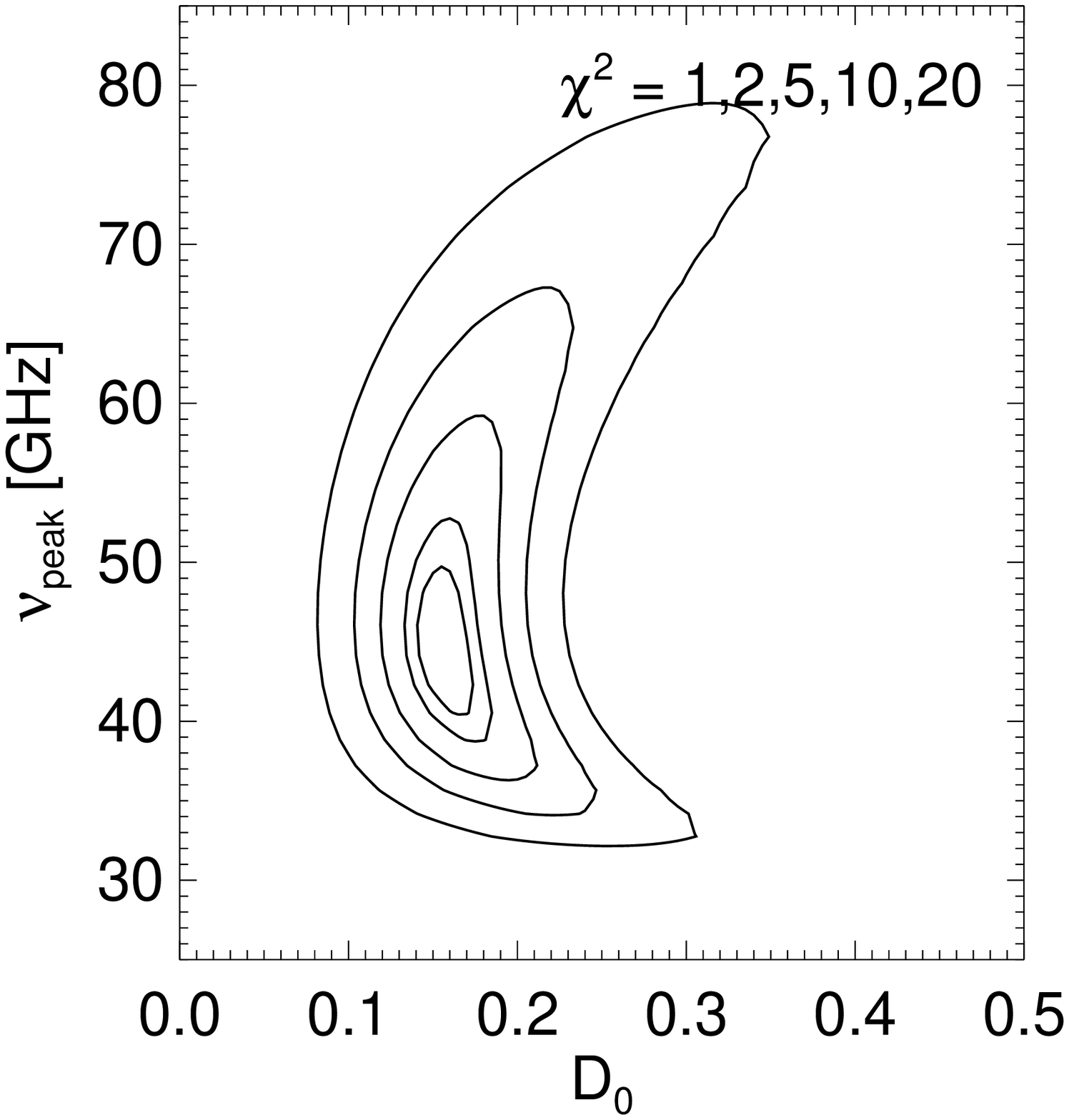}
  \includegraphics[width=0.23\textwidth]{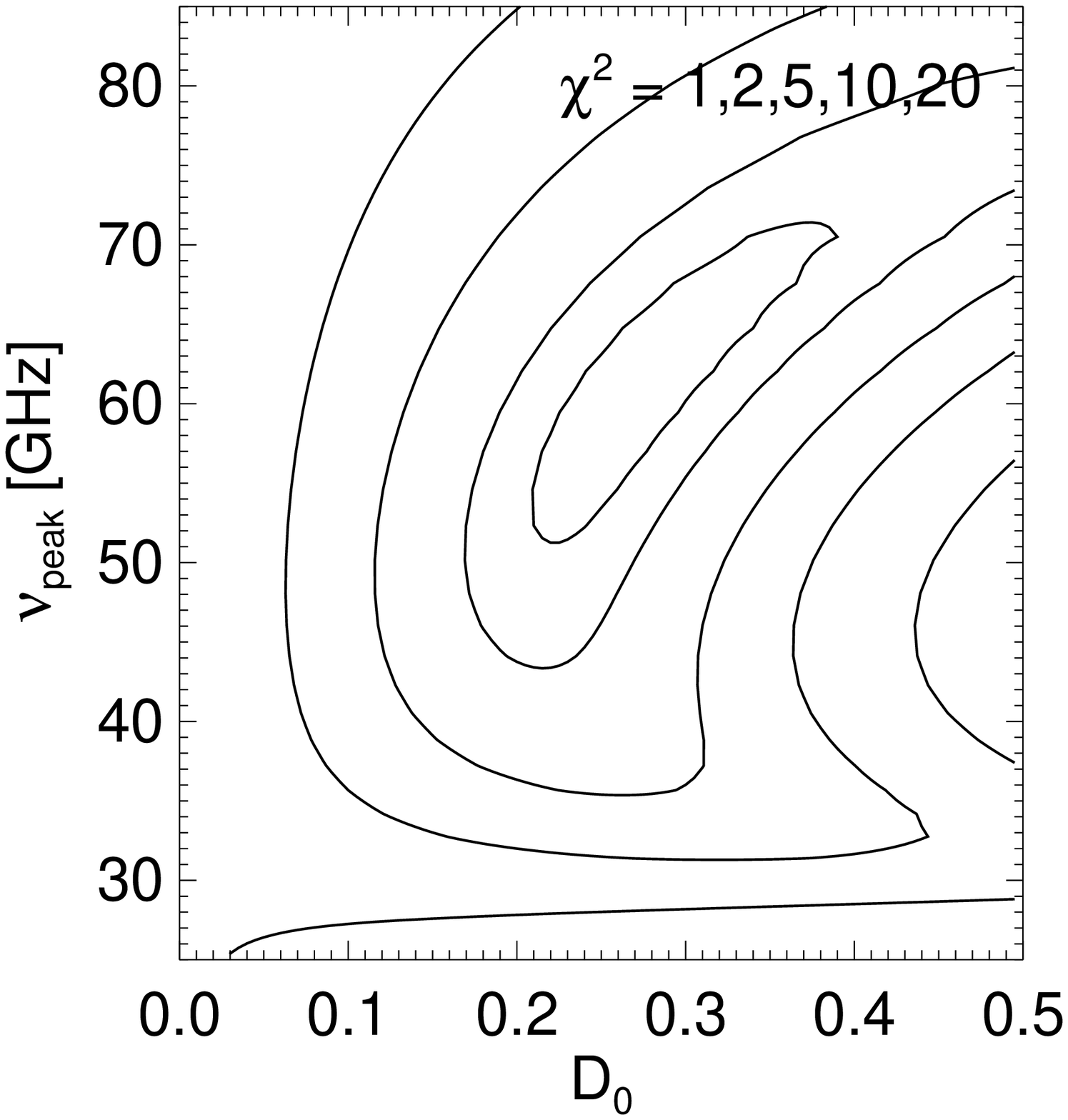}
  \includegraphics[width=0.23\textwidth]{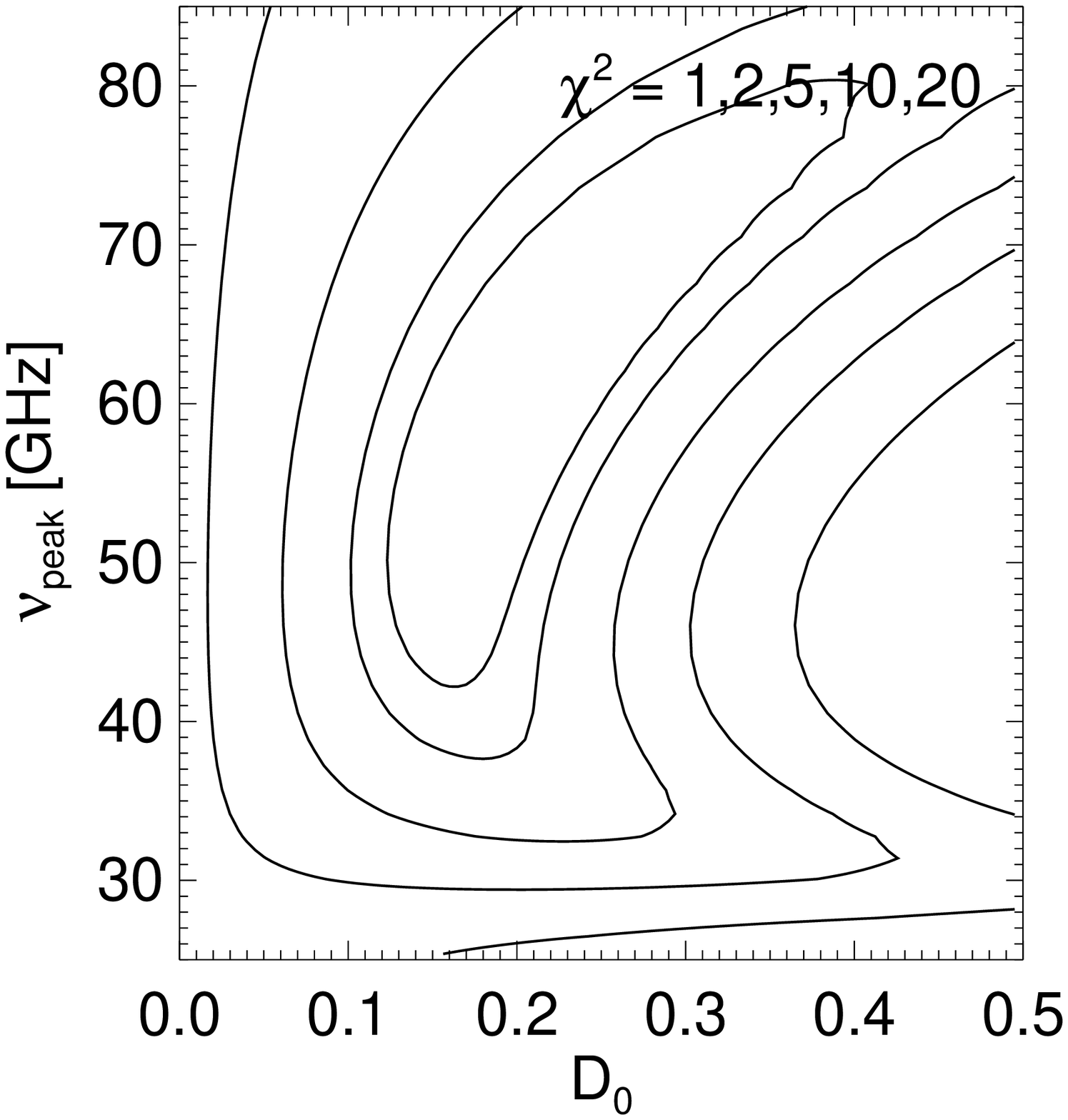}
}
\caption{
The same as \reffig{hdc-fs} except for the four regions of the Gum Nebula (see 
\reftbl{fit-res}).  Although the peak frequency is only loosely constrained, the 
$D_0=0$ null hypothesis is ruled out at $\geq 12\sigma$ in all regions.
}\label{fig:hdc-rg}
\epm

\section{Spectrum of \hal-correlated emission}
\label{sec:spectra}

As noted above, in DF07 we showed that the spectrum of \hal-correlated 
emission in the WMAP data has a bump around $\nu \sim 50$ GHz that we now argue 
is consistent with a spinning dust contamination of the familiar \refeq{ff-spec} 
free--free spectrum.  We also point out in DF07 that any measurement of a 
foreground spectrum is necessarily contaminated by a component which has the 
frequency dependence of the CMB due to imperfect foreground removal when 
constructing a CMB template.\footnote{Throughout this paper, we will work with 
the results found using our CMB5 estimator for the CMB -- see DF07 for details.}  
However, as we will show shortly, this contamination is incapable of producing 
the observed bump.

\reffig{hdc-fs} shows our full sky fit for \hal-correlated emission.  Our 
hypothesis is that this spectrum can be well fit by a linear combination of 
free--free, spinning dust, and CMB spectra.  Thus, our model spectrum is
\bel{mod}
  I_{\nu}^{\rm mod} = F + D + C,
\ee
where
\bel{ff-comp}
  F = F_0 \left(\frac{\nu}{23\mbox{ GHz}}\right)^{-0.15},
\ee
\bel{sd-comp}
  D = D_0 \times \mbox{ (DL98 WIM model with $\nu_{p} = 50$ GHz)},
\ee
and
\bel{cmb-comp}
  C = C_0 \left(\frac{\nu}{23\mbox{ GHz}}\right)^2.
\ee
We minimize $\chi^2 = \sum_{i} (I_{\nu_i}^{\rm mod} - I_{\nu_i})^2/\sigma_i^2$ 
over the parameters $F_0$, $D_0$, and $C_0$ which are scaled to be of order 
unity.  Here $\sigma_i$ are the formal errors in band $i$ of the spectrum from 
our DF07 template fit, and we have one degree of freedom.

As shown in \reffig{hdc-fs}, the spectral model in \refeq{mod} fits the data 
astonishingly well.  The free--free and spinning dust amplitudes are comparable 
while the contamination from the CMB is minimal.  The very low value of 
$\chi^2=1.84$ reflects the quality of the fit. \reffig{hdc-fs} also shows 
$\Delta\chi^2=$ 1,2,3,4, and 5 contours in the $D_0$-$C_0$ plane minimizing 
$\chi^2$ over $F_0$ at each point, and holding $\nup$ fixed at 50 GHz.  The 
null hypothesis that only a free--free plus CMB spectrum (i.e., $D_0 = 0$) fit 
the data is ruled out to a very high significance.  Thus, either free--free 
emission does not follow \refeq{ff-spec}, or the \hal\ map is tracing
an additional emission component.  This other emission component seems
to be well fit by a shifted WIM spinning dust spectrum.

Lastly, the $\chi^2$ contours in the $D_0$-$\nu_{\rm peak}$ plane in the bottom 
panel of \reffig{hdc-fs} indicate that, although the peak frequency of the 
spinning dust is not very well constrained, values of $\sim 45-55$ GHz are 
consistent with the data.  Peak frequencies less than 40 GHz and greater than 
60 GHz are ruled out at $4\sigma$.

\bp
\centerline{
  \includegraphics[width=0.23\textwidth]{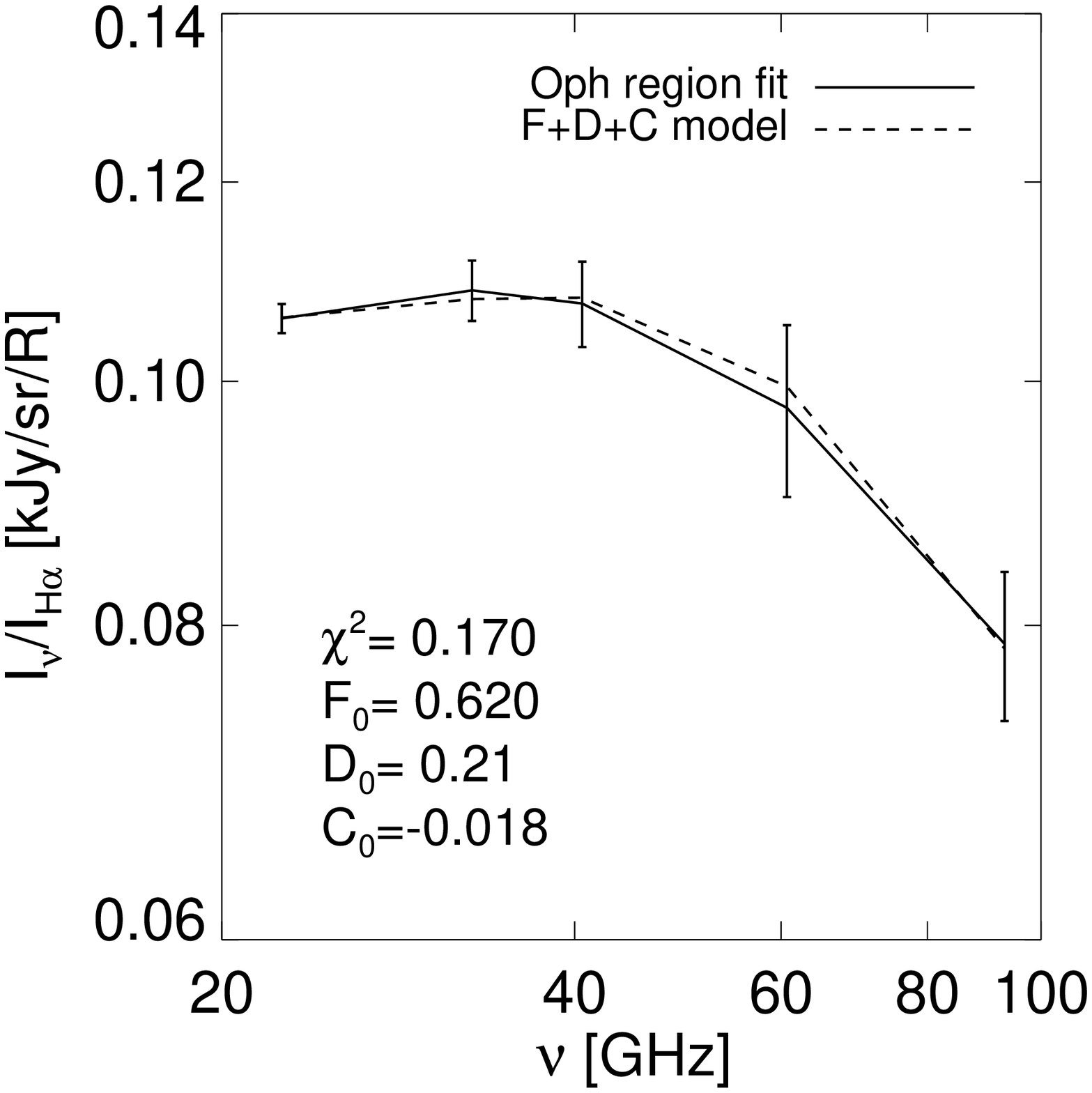}
  \includegraphics[width=0.23\textwidth]{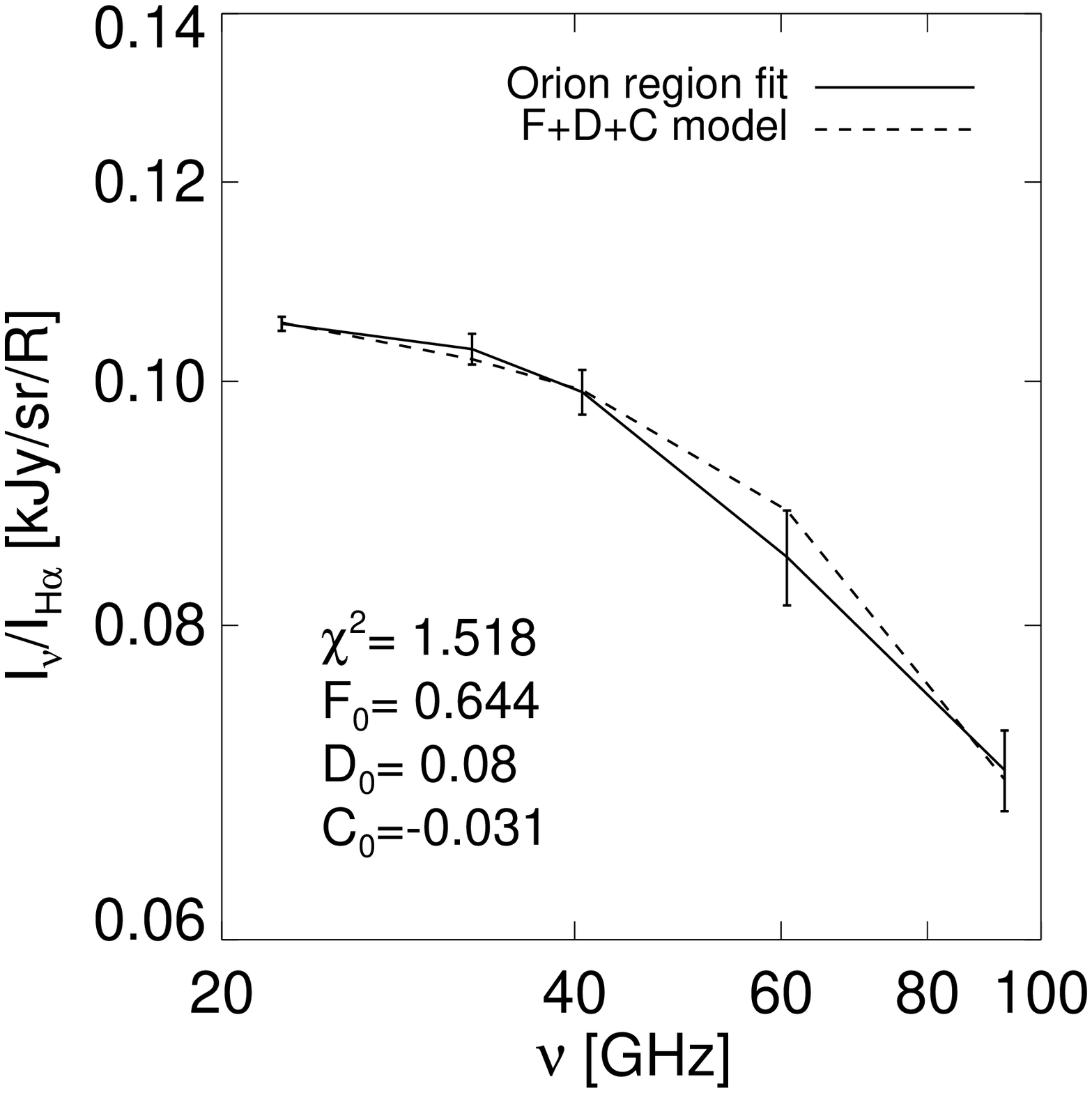}
}
\centerline{
  \includegraphics[width=0.23\textwidth]{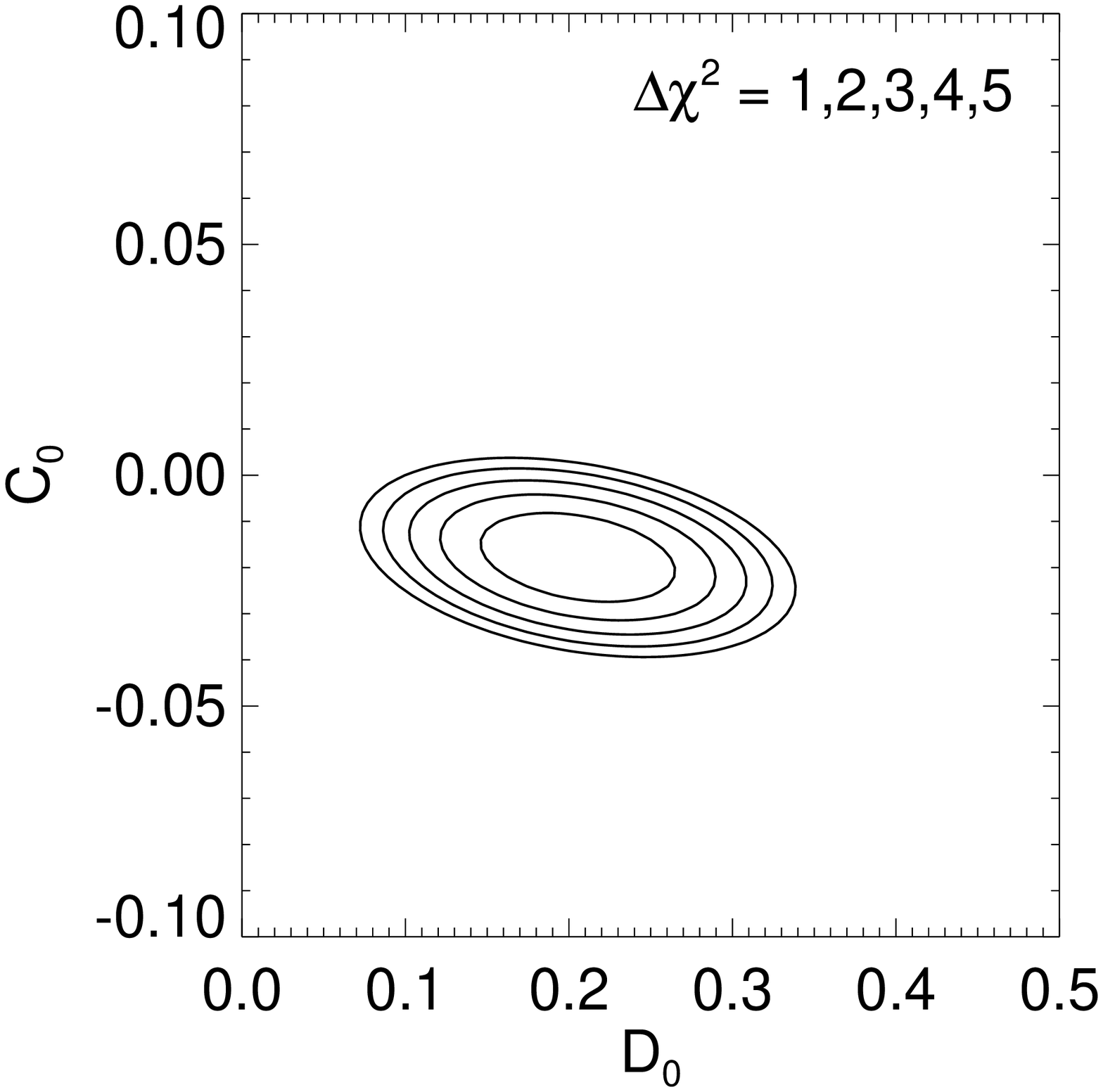}
  \includegraphics[width=0.23\textwidth]{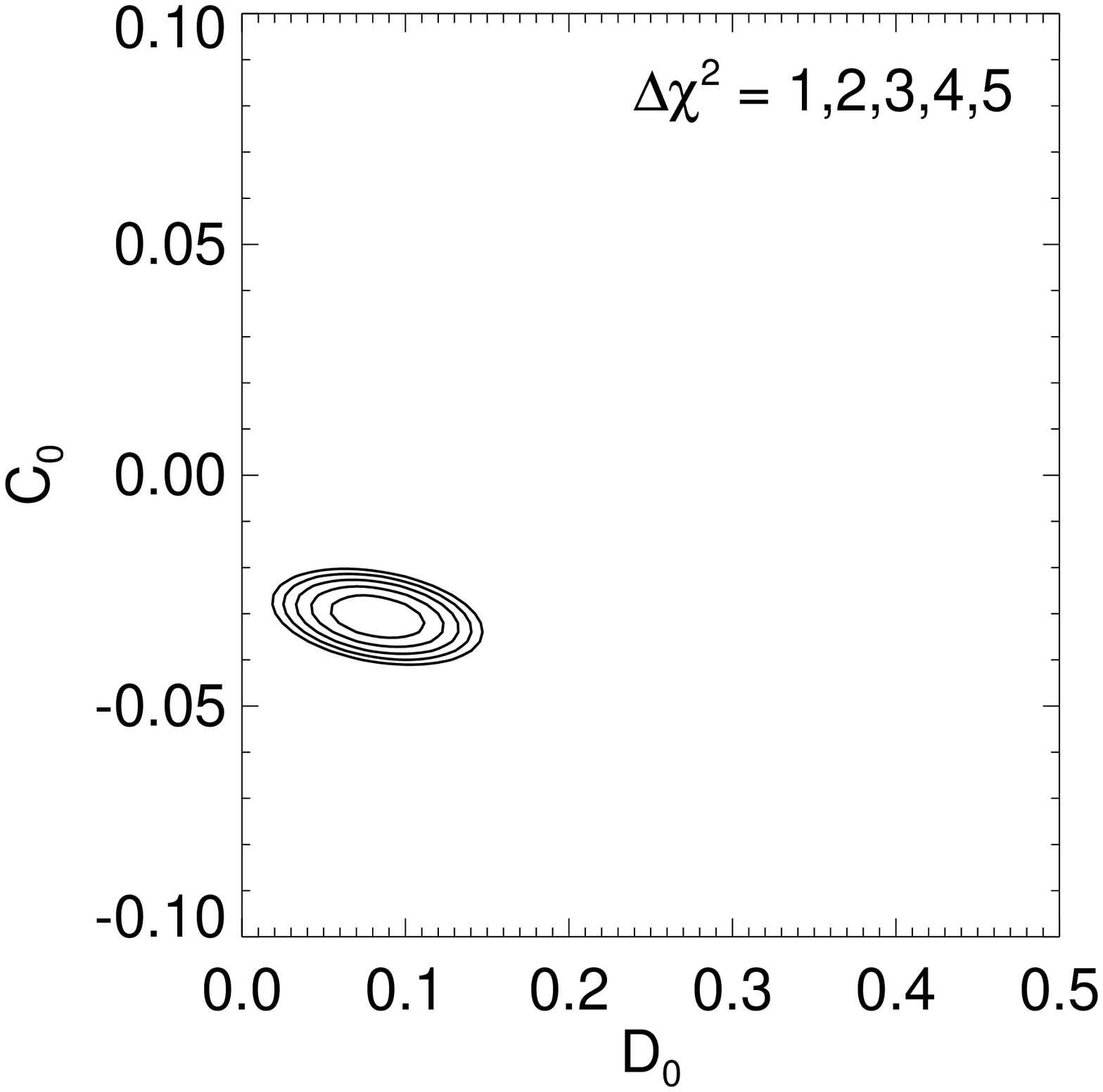}
}
\centerline{
  \includegraphics[width=0.23\textwidth]{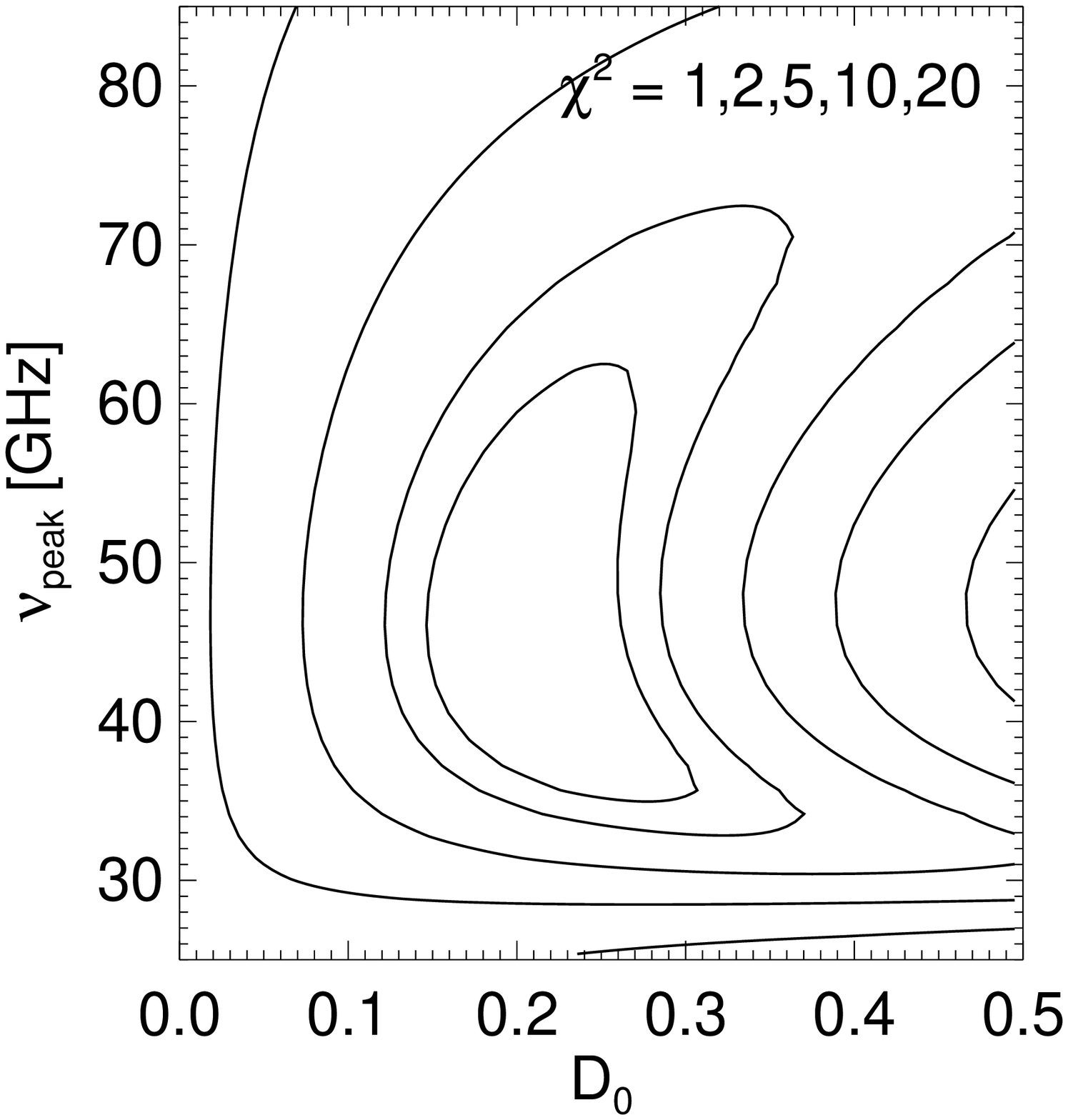}
  \includegraphics[width=0.23\textwidth]{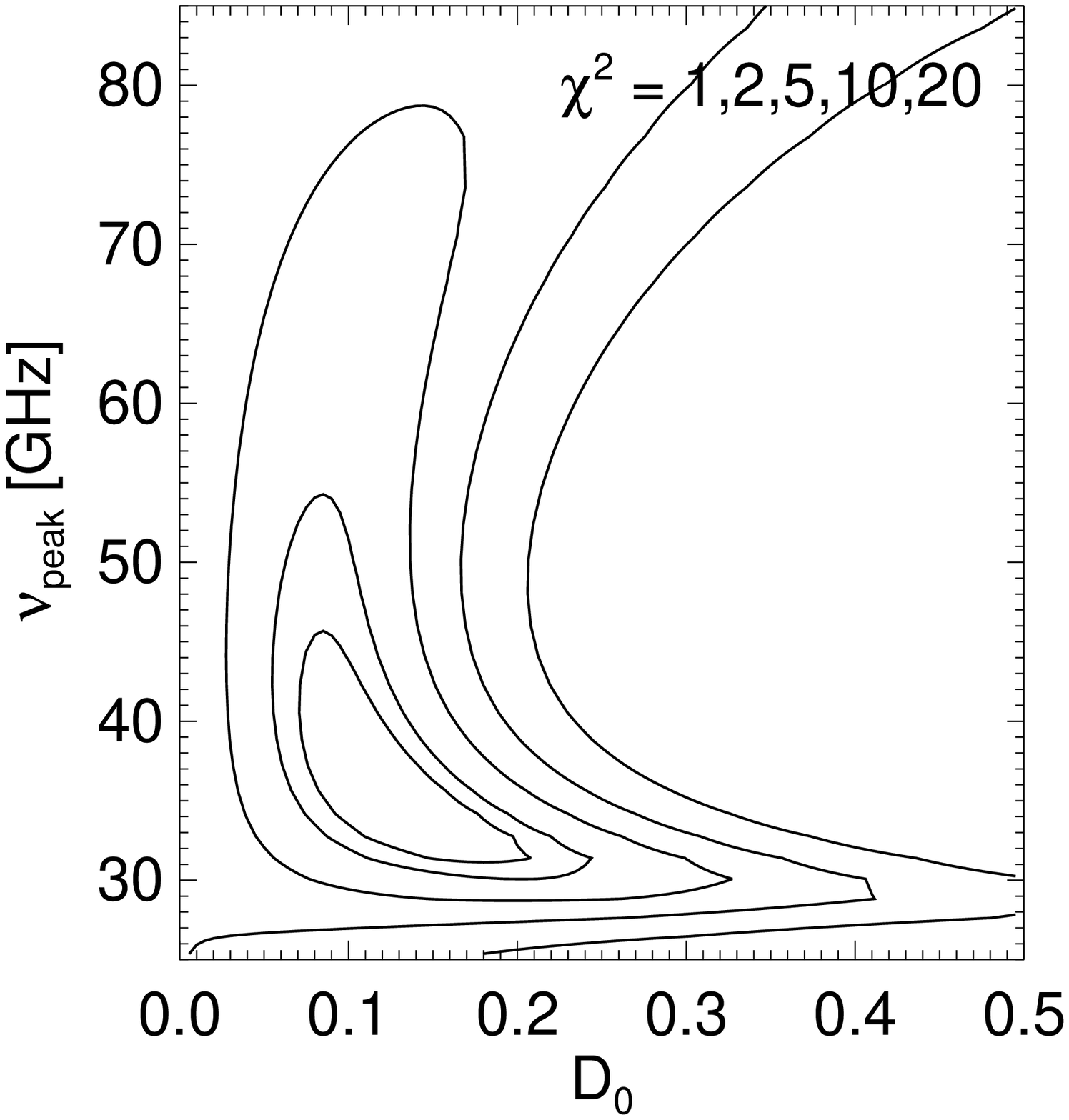}
}
\caption{
The same as \reffig{hdc-fs} except for a region which encloses the Ophiuchus 
Complex (left column) and one which is near the Orion Nebula (right column).  
See \reftbl{fit-res} for details.
}\label{fig:hdc-op-or}
\ep

\begin{deluxetable*}{|c|c|c|c|c|c|c|c|}
\tablehead{
 & Full sky & \multicolumn{4}{c|}{Gum Nebula} & Ophiuchus & Orion \\
 & & 1 & 2 & 3 & 4 & &
}
\startdata
 lrange & 0:360 & 235:256 & 256:280 & 235:256 & 256:280 & 334:13 & 195:230 \\ 
 \hline
 brange & -90:90 & -30:0 & -30:0 & 0:30 & 0:30 & 0:90 & -35:0 \\ 
 \hline
 $\chi^2$ & 1.84 & 0.17 & 1.10 & 1.08 & 0.26 & 0.17 & 1.52 \\ 
 \hline
 $\Delta\chi^2_{D=0}$ & 335.02 & 43.84 & 87.32 & 17.59 & 12.09 & 11.89 & 8.30 \\ 
 \hline
 $ I_{F}/I_{D} (\nu=\nu_{\rm peak}) $ & 6.23 & 4.70 & 6.07 & 5.38 & 6.80 & 4.75 & 12.22
\enddata
\tablecomments{
Region definitions (in Galactic coordinates $l$ and $b$) and results of the fits 
presented in Figures \ref{fig:hdc-fs}-\ref{fig:hdc-op-or}.  The very low values 
of $\chi^2$ reflect both the quality of the fits and the small number of 
constraints (number of degrees of freedom = 2).  The $D_0=0$ null hypothesis is 
ruled out at $> 8\sigma$ in \emph{all} regions, while the relative intensity of 
free--free to spinning dust emission at $\nu_{\rm peak} = 50$ GHz is in the 
range 4.7-12.2.
}\label{tbl:fit-res}
\end{deluxetable*}

\bpm
\centerline{
  \includegraphics[width=0.75\textwidth]{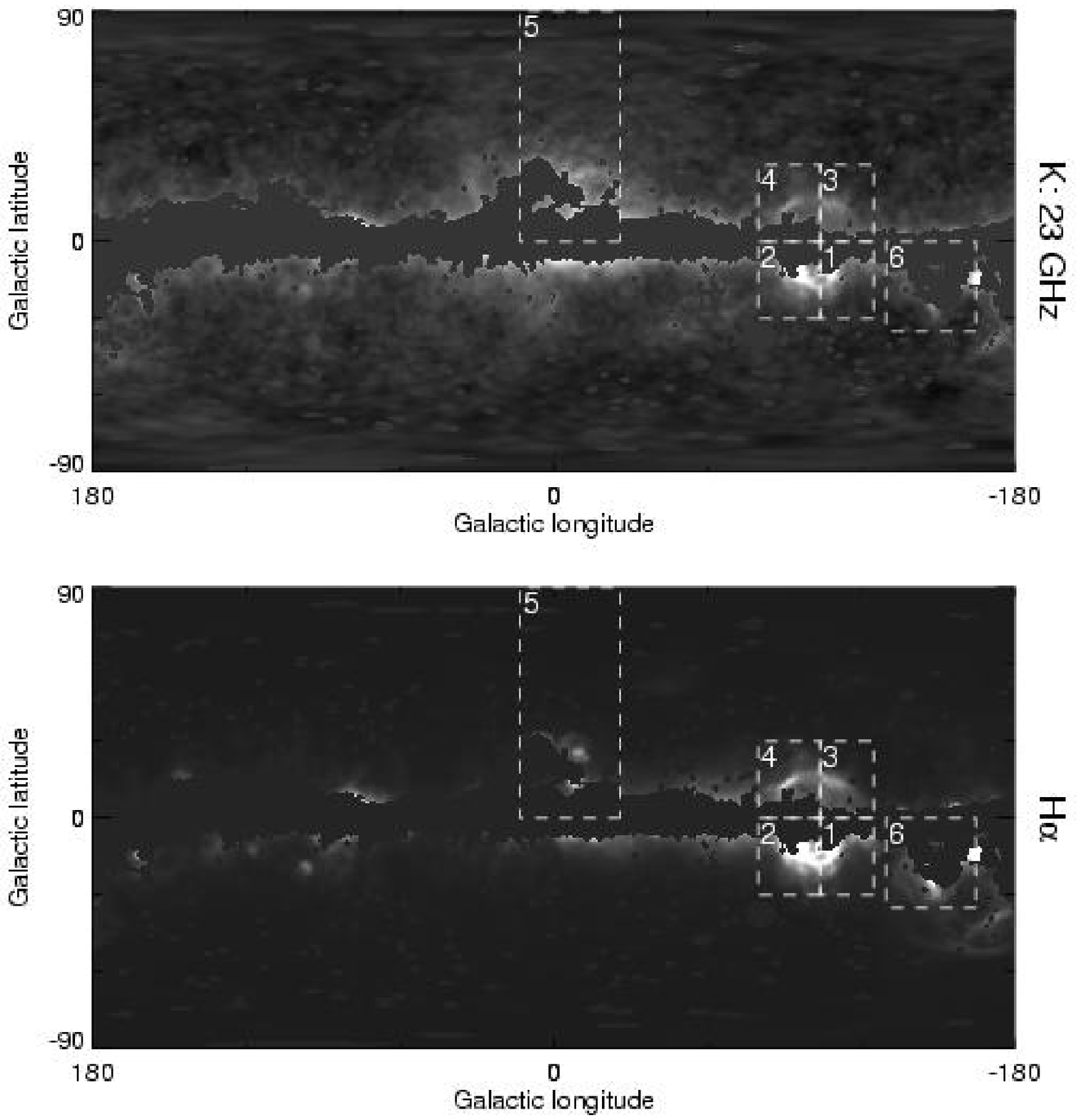}
}
\caption{
WMAP K band data (\emph{top}) and the \hal\ map (\emph{bottom}) with our different 
fit regions shown as dotted boxes.  In addition to the full sky fit, the regions 
are the Gum Nebula (1,2,3,4), the Ophiuchus complex (5), and the Orion Nebula (6).  
See \reftbl{fit-res} for details.
}\label{fig:regions}
\epm

The next step is to consider smaller individual regions of the sky to show that 
the signal persists.  Figures \ref{fig:hdc-rg} and \ref{fig:hdc-op-or} show fits 
for four different regions of the Gum Nebula, the north central Galactic sky 
(including the Ophiuchus complex), and a region near the Orion Nebula (see 
\reftbl{fit-res} and \reffig{regions} for details). Though the fits are 
significantly noisier as a 
result of the smaller number of pixels used, the bump in 
the \hal-correlated emission remains in all regions, with the free--free
amplitude roughly 4-7 
times larger than the spinning dust.  Furthermore, the case of no spinning dust 
$D_0 = 0$ is ruled out in \emph{all} regions at $\geq 8\sigma$.  Although the 
$\chi^2$ contours in the $D_0$-$\nu_{\rm peak}$ plane offer a tantalizing hint 
that the peak frequency of the spinning dust is lower in the Orion Nebula region 
than the others, the present data are simply too noisy to make a statistically 
significant statement.

\begin{deluxetable}{|c|c|c|c|c|}
\tablehead{
   & H$\alpha$ & Dust & Haslam & Haze
}
\startdata
  H$\alpha$ &  1.00 &  0.45 &  0.18 &  0.10 \\
  Dust      &  0.45 &  1.00 &  0.56 &  0.26 \\
  Haslam    &  0.18 &  0.56 &  1.00 &  0.58 \\
  Haze      &  0.10 &  0.26 &  0.58 &  1.00
\enddata
\tablecomments{
The correlation matrix for the templates presented in \citet{DF07}.  These are 
the templates that were used in the multi-linear regression fit which revealed 
the bump in the \hal-correlated emission presented in Figures 
\ref{fig:hdc-fs}-\ref{fig:hdc-op-or}.
}\label{tbl:corrmat}
\end{deluxetable}

Lastly, we address the possibility that the bump in the \hal-correlated 
emission represents a cross-correlation bias between the different templates 
used in DF07.  That is to say, perhaps the \hal\ spectrum is absorbing power from 
other templates in the fit.  Table \ref{tbl:corrmat} shows the correlation 
matrix, defined as
\bel{corrmat}
  \Gamma_{i,j} = \frac{\langle T_i T_j \rangle}
								{\sqrt{\langle T_i^2 \rangle\langle T_j^2 \rangle}}
\ee
where $\Gamma_{i,j}$ is the correlation between the $i$th and $j$th template.  
The correlation between the templates is rather small.  In particular, the dust 
map is the template which is most highly correlated with the \hal\ map, and even 
then, the variance in the \hal-correlated emission can be at most 
$\Gamma_{\mbox{\hal, dust}}^2 = 20\%$ due to correlation with dust.

\bp
\centerline{
  \includegraphics[width=0.24\textwidth]{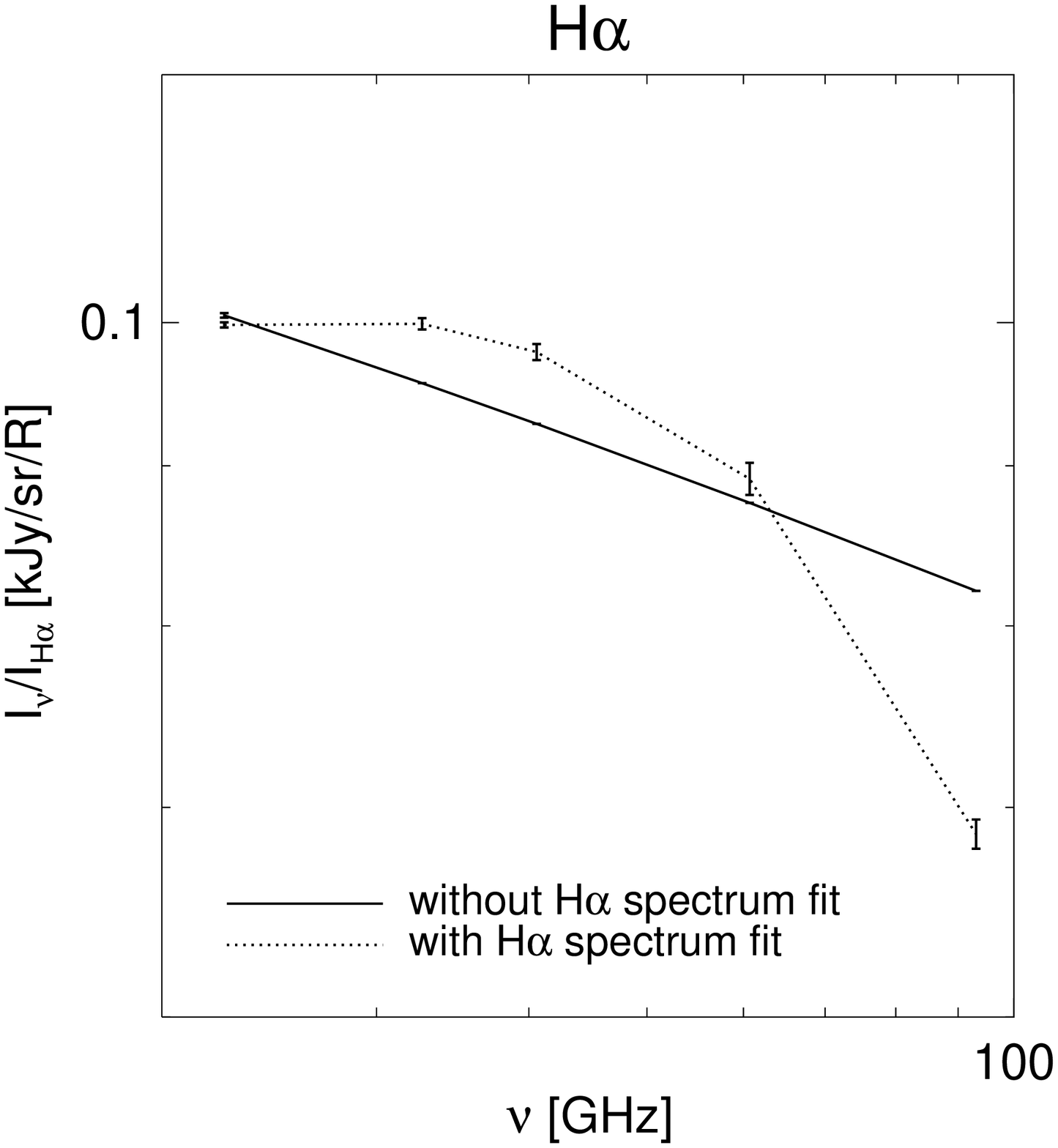}
  \includegraphics[width=0.24\textwidth]{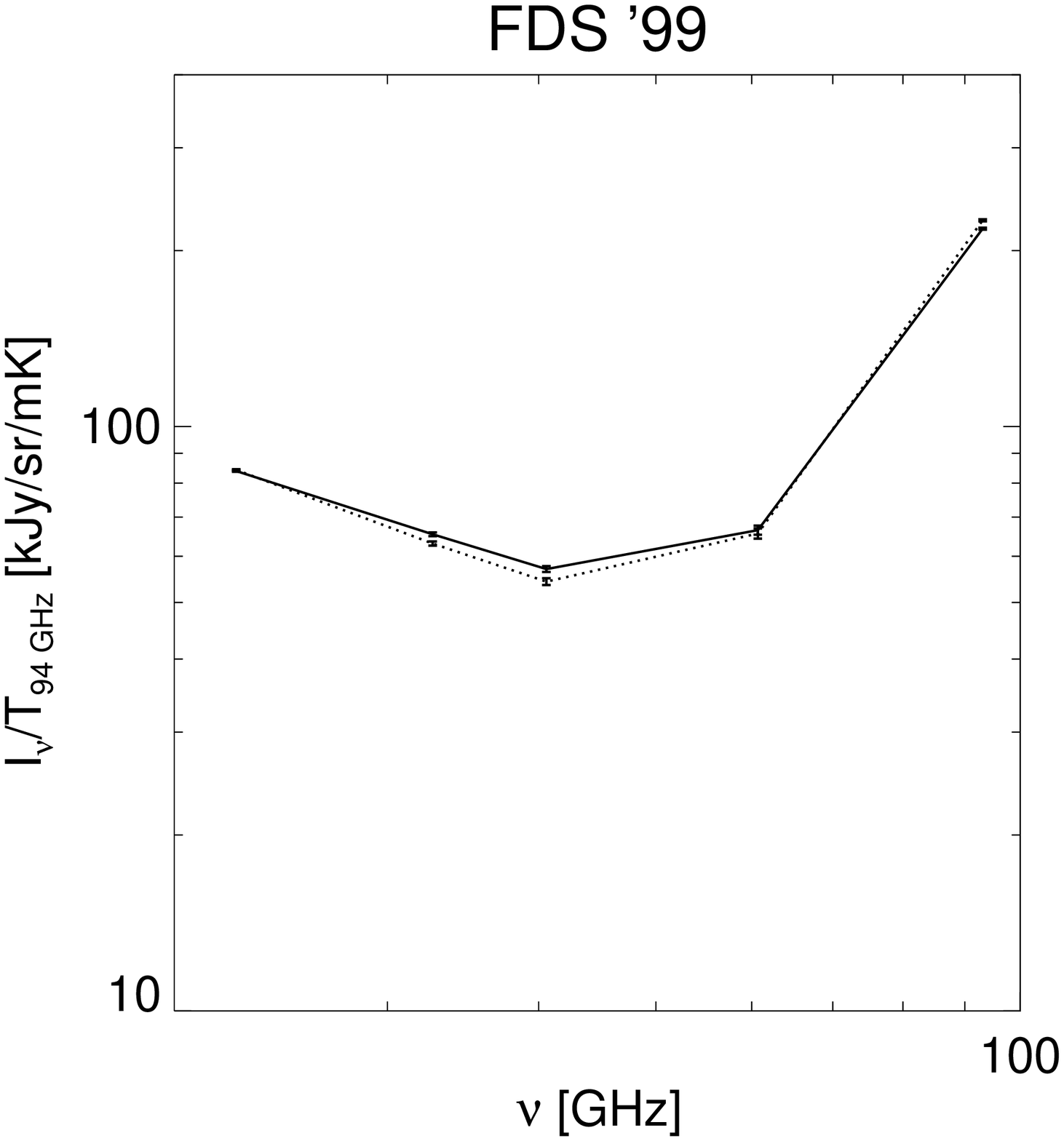}
}
\centerline{
  \includegraphics[width=0.24\textwidth]{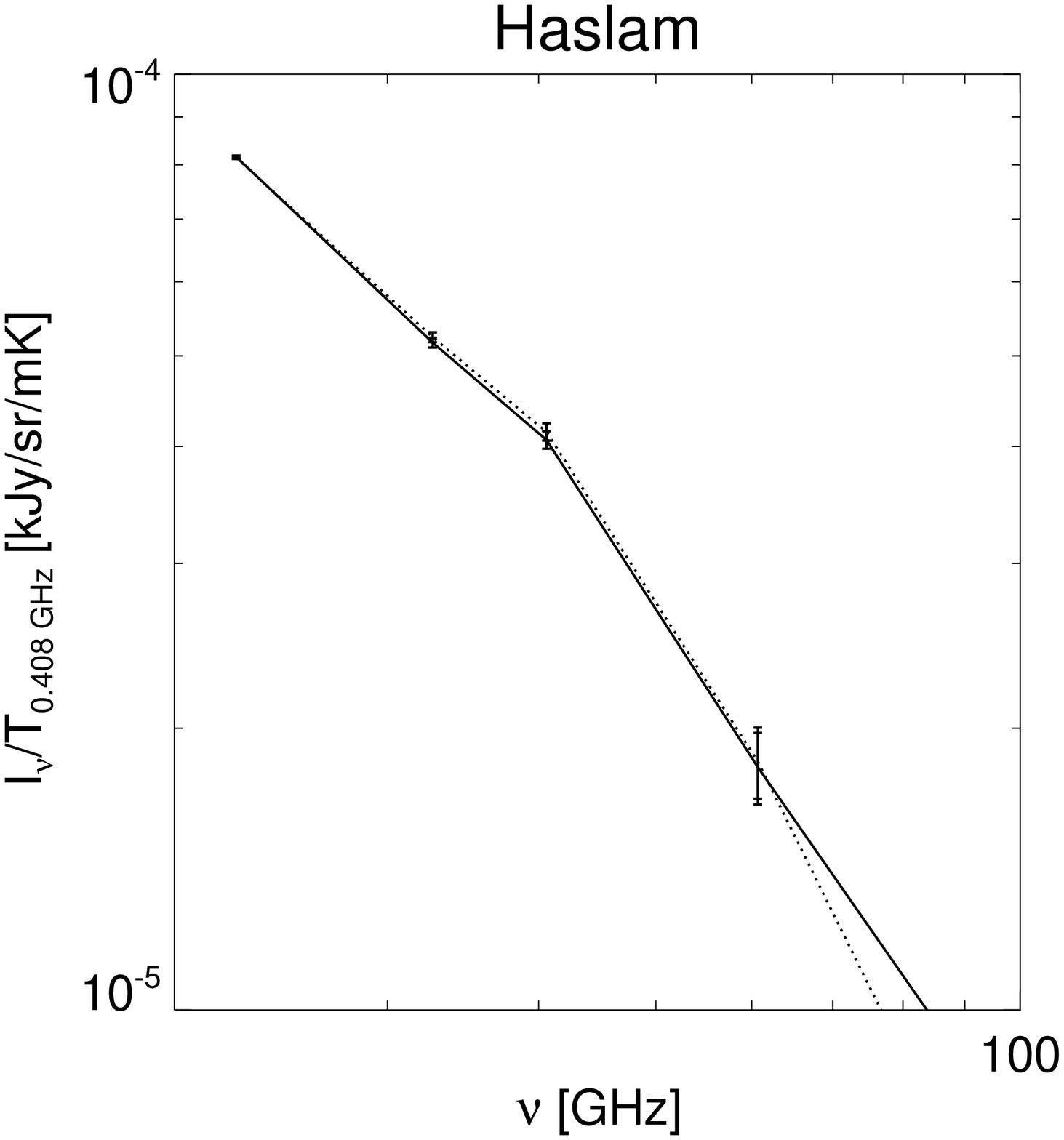}
  \includegraphics[width=0.24\textwidth]{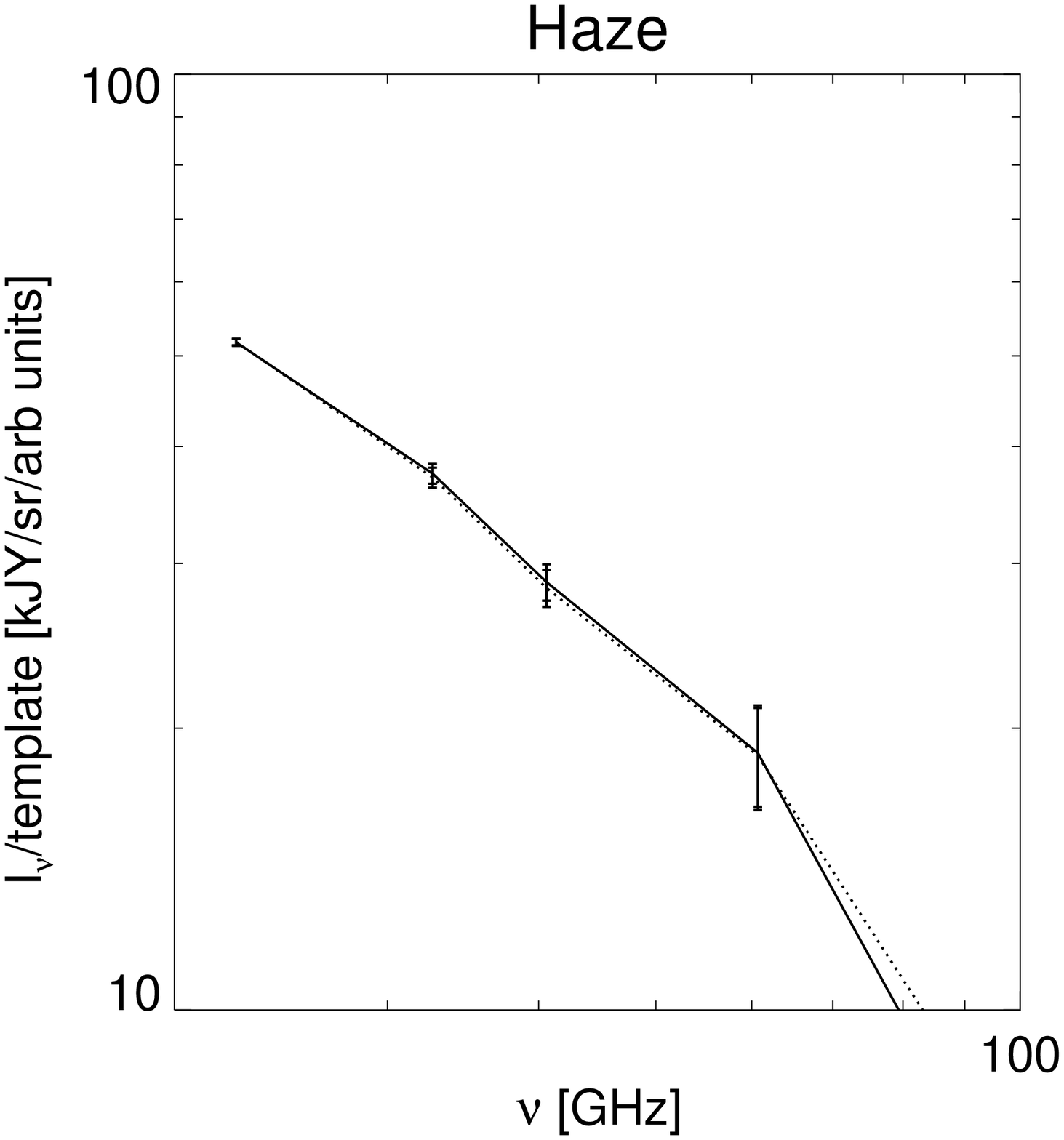}
}
\caption{
The intensity of the emission correlated with the four templates presented in 
\citet{DF07}.  The solid lines indicate fits for which the \hal-correlated 
emission was constrained to follow a $I_{ff} \propto \nu^{-0.15}$ (see 
\refeq{ff-spec}) frequency dependence, while the dotted lines indicate that the 
\hal-correlated spectrum was fit explicitly.  Both the insensitivity of the 
inferred dust, Haslam, and haze -correlated spectra to fixing the \hal\ spectrum 
as well as the relatively low correlation among the templates (\reftbl{corrmat}) 
indicates that the spinning dust bump is indeed correlated primarily with \hal\ 
emission.
}\label{fig:corr-hal}
\ep

In fact, \reffig{corr-hal} shows that the effects are much smaller.  Here, the 
solid lines show the resultant spectra for all foreground components both with 
and without fitting the \hal\ spectrum.  In the latter case the \hal\ emission was 
assumed to follow the free--free spectrum in \refeq{ff-spec}.  The features of 
these spectra are discussed at length in DF07, but the important point is that 
the other foregrounds are not significantly affected when the \hal\ spectrum is 
fixed.  This is strong evidence that the emission mechanism causing the bump in 
the \hal\ spectrum is highly correlated with the \hal\ template.

\section{Discussion}
\label{sec:discussion}

In \citet{DF07} we identified a deviation from the classical
free--free (thermal bremsstrahlung) spectrum in the \hal-correlated
emission of the 3-year WMAP data.  In this paper we have argued that
the spectral data are consistent with a superposition of free--free
and spinning dust emission from a ``warm ionized medium'' \citep[WIM;
see][]{DL98b}.  Additionally, there is contamination by a component with the 
spectrum of the CMB due to chance morphological correlation between the CMB and 
the \hal\ map \citep[see][for details]{DF07}.

Since the intensity of both free--free and WIM spinning dust emission scale 
proportionately with the number density squared (free--free emission is 
generated by the collision of free electrons with ions while the WIM spinning 
dust emission comes from collision of ions with tiny dust grains), both should 
be traced by an emission measure map like the \hal\ map.

We have assumed throughout that the dust/gas ratio is constant and the
grain size distribution is the same everywhere, so that $I \propto \int
n_i n_d dl \propto \int n_i^2$. 

We have studied both the (nearly) full sky and smaller independent regions of 
interest including the Gum Nebula, the Oph complex, and the Orion Nebula.  
For all regions, the \hal-correlated emission is well fit by a linear 
combination of an $I_{\nu} \propto \nu^{-0.15}$ power law for free--free and a 
\citet{DL98b} WIM spinning dust model shifted in frequency by a factor of 2 and 
with only a minimal adjustment in amplitude.  Specifically, the spinning 
dust component is $\sim$ 4-7 times weaker than the free--free component.

The null hypothesis that a free--free only spectrum is consistent with the data 
is ruled out at $\geq 8\sigma$ confidence for all regions considered here, 
despite the substantial noise in the data.  Furthermore, we have shown that 
the bump in the \hal-correlated emission spectrum is \emph{not} due to chance 
cross correlation between the \hal\ map and the additional foreground templates 
presented in \citet{DF07}, indicating that it is indeed correlated primarily 
with \hal\ emission.

\citet{F02} listed a number of criteria by which the existence of
spinning dust could be established.  Among them were that the peak
frequency must shift as a function of the environment in a way at
least qualitatively consistent with DL98.  In this work we find in the
WMAP data a WIM-correlated component with a peak frequency of 50 GHz,
about a factor of two higher than the peak frequency of the spinning 
dust component that is spatially correlated with thermal dust emission (i.e. WNM 
or CNM spinning dust).  That this
frequency is higher is in accordance with the DL98 models, strongly
supporting the spinning dust hypothesis.  However, because the shift
is larger than expected, a careful assessment of the situation
requires that the reason for the size of the shift be determined.  If,
for example, the WIM dust has smaller grains, then a correlated
measurement of $R_V\equiv A_V/E_{B-V}$ using e.g. background stars
would complete the case for spinning dust. 

In summary, there is clearly an additional foreground component
correlated with \hal\ emission and its spectral shape is consistent
with spinning dust, though at a somewhat higher frequency than
expected.  However, the increased sensitivity, frequency coverage, and
finer spatial resolution of future missions like \emph{Planck} may provide the
necessary evidence.

We acknowledge helpful discussions with Bruce Draine, Gary Hinshaw,
Joanna Dunkley, et al.  Some of the results in this paper were derived
using HEALPix\footnote{The HEALPix web page is 
\texttt{http://healpix.jpl.nasa.gov}} \citep{gorski05}.
This research made use of the NASA Astrophysics Data System (ADS) and
the IDL Astronomy User's Library at Goddard\footnote{Available at
\texttt{http://idlastro.gsfc.nasa.gov}}.  DPF and GD are supported in
part by NASA LTSA grant NAG5-12972.

\end{document}